\documentclass[11pt]{article}
\usepackage{fixltx2e,fix-cm}
\usepackage{amssymb}
\usepackage{amsmath}
\usepackage{graphicx}
\usepackage{subfigure}
\usepackage{makeidx}
\usepackage{multicol}
\usepackage{mathptmx}
\usepackage[T1]{fontenc}
\usepackage[utf8]{inputenc}
\usepackage{authblk}

\input epsf             

\def\qed{\hbox{$\square$}}

\newenvironment{gs:proof}{\noindent\textit{Proof.}\quad}{\hfill\qed\medskip}

\newtheorem{thm}{Theorem}[section]

\newtheorem{example}[thm]{Example}
\newtheorem{definition}[thm]{Definition}
\def\qed{\hfill\ensuremath{\blacksquare}}

\newcommand\beq{\begin{equation}}
\newcommand\eeq{\end{equation}}
\newcommand\bce{\begin{center}}
\newcommand\ece{\end{center}}
\newcommand\bea{\begin{eqnarray}}
\newcommand\eea{\end{eqnarray}}
\newcommand\ba{\begin{array}}
\newcommand\ea{\end{array}}
\newcommand\ben{\begin{enumerate}}
\newcommand\een{\end{enumerate}}
\newcommand\bit{\begin{itemize}}
\newcommand\eit{\end{itemize}}
\newcommand\brr{\begin{array}}
\newcommand\err{\end{array}}
\newcommand\bt{\begin{tabular}}
\newcommand\et{\end{tabular}}

\def\petitcarre{\vrule height4pt width 4pt depth0pt}
\def\QED{\relax\ifmmode\eqno{\hbox{\petitcarre}}\else{%
  \unskip\nobreak\hfil\penalty50\hskip2em\hbox{}\nobreak\hfil
  \petitcarre
  \parfillskip=0pt \finalhyphendemerits=0\par\smallskip}
  \fi}

\newcommand{\paper}{chapter}

\title{Tree-decomposable and Underconstrained Geometric	Constraint Problems}
\author[1]{Ioannis Fudos\thanks{fudos@cse.uoi.gr}}
\author[2]{Christoph M. Hoffmann\thanks{cmh@purdue.edu}}
\author[3]{Robert Joan-Arinyo\thanks{robert@cs.upc.edu}}
\affil[1]{Department of Computer Science and Engineering, University of Ioannina, Ioannina, Greece}
\affil[2]{Department of Computer Science, Purdue University, West Lafayette, IN, USA}
\affil[3]{Department of Computer Science, Universitat Politecnica de Catalunya, Barcelona, Catalonia}

\begin{document}
\bibliographystyle{plain}

\maketitle

\begin{abstract}
	In this paper, we are concerned with geometric constraint solvers, i.e., with programs that find one or more solutions of a geometric constraint problem. If no solution exists, the solver is expected to announce that no solution has been found. Owing to the complexity, type or difficulty of a constraint problem, it is possible that the solver does not find a solution even though one may exist. Thus, there may be false negatives, but there should never be false positives. Intuitively, the ability to find solutions can be considered a measure of solver's competence.
	
	We consider static constraint problems and their solvers. We do not consider dynamic constraint solvers, also known as dynamic geometry programs, in which specific geometric elements are moved, interactively or along prescribed trajectories, while continually maintaining all stipulated constraints. However, if we have a solver for static constraint problems that is sufficiently fast and competent, we can build a dynamic geometry program from it by solving the static problem for a sufficiently dense sampling of the trajectory of the moving element(s).
	
	The work we survey has its roots in applications, especially in mechanical computer-aided design (MCAD). The constraint solvers used in MCAD took a quantum leap in the 1990s. These approaches solve a geometric constraint problem by an initial, graph-based structural analysis that extracts generic subproblems and determines how they would combine to form a complete solution. These subproblems are then handed to an algebraic solver that solves the specific instances of the generic subproblems and combines them.
	
\end{abstract}
%
%
%


\hyphenation{geo-metry}

\hyphenation{para-meter}




\section{Introduction, Concepts and Scope}
\label{sec:intro}

A geometric constraint problem, also known as a geometric constraint
system, consists of a finite set of geometric objects, such as points,
lines, circles, planes, spheres, etc., and constraints upon them, such
as incidence, distance, tangency, and so on.  A solution of a
geometric constraint problem {\em P} is a coordinate assignment for
each of the geometric objects of {\em P} that places them in relation
to each other such that all constraints of {\em P} are satisfied.  A
problem {\em P} may have a unique solution, it may have more than one
solution, or it may have no solution.

In this \paper, we are concerned with geometric
constraint solvers, i.e., with programs that find one or more solutions of a geometric
constraint problem.  If no solution exists, the solver is expected to
announce that no solution has been found.  Owing to the complexity,
type or difficulty of a constraint problem, it is possible that the
solver does not find a solution even though one may exist.  Thus,
there may be false negatives, but there should never be false
positives.  Intuitively, the ability to find solutions can be
considered a measure of solver's competence.

We consider static constraint problems and their solvers.  We do not
consider dynamic constraint solvers, also known as dynamic geometry
programs, in which specific geometric elements are moved,
interactively or along prescribed trajectories, while continually
maintaining all stipulated constraints.  However, if we have a solver
for static constraint problems that is sufficiently fast and
competent, we can build a dynamic geometry program from it by solving
the static problem for a sufficiently dense sampling of the trajectory
of the moving element(s).

The work we survey has its roots in applications, especially in mechanical computer-aided design (MCAD).
The constraint solvers used in MCAD took a quantum leap with the work by Owen \cite{bib:owen91}.
Owen's algorithm solves a geometric constraint problem by an initial, graph-based structural analysis that
extracts generic subproblems and determines how they would combine to form a complete solution.
These subproblems are then handed to an algebraic solver that solves the specific instances of
the generic subproblems and combines them.
Owen's graph analysis is top down.  A bottom-up analysis was proposed in \cite{bib:bouma95}.
Subsequent work expanded the knowledge of
\begin{enumerate}
	\item
the structure and properties of the constraint graph, see Section \ref{sec:CGraph};
\item
the geometric vocabulary, see Section \ref{sec:CSolver};
\item
the understanding of spatial constraint systems, see Section \ref{sec:C3d}.	
\end{enumerate}
We also look briefly at under-constrained problems in Section~\ref{sec:undercon}.

Restricted to points and distances, the constraint graph analysis has deep
roots in mathematics and combinatorics; see, e.g., \cite{bib:Maxwell64,bib:Henneberg08,bib:laman70}
for some of these connections.
Here, we limit the discussion to constraint problems that are motivated by MCAD applications, and that means that the geometric vocabulary has to be richer than points and distances between them.
In the remainder of this section we introduce informally major concepts and methods
for solving geometric constraint systems (GCS) with an application perspective in mind. 

Note in particular that some definitions in this section may be {\em provisional}.
Those definitions, although formally incorrect, are so given nevertheless because
they facilitate understanding the material.
They are clearly identified along with examples that show where they fall short ---
and what can be done about it.

\subsection{Geometric Constraint Systems (GCS)}

Fix the space in which to consider a geometric constraint system
(GCS).  Typical examples are the Euclidean space in 2 or 3 dimensions.
Each object to be placed by a GCS instance in that space has a
specific number of degrees of freedom (dof), i.e., a specific number
of independent coordinates.  In Euclidean 2-space, points and lines
each have 2 dof.  In Euclidean 3-space, a point has 3 dof and a line
has 4.  A constraint on such objects corresponds to one or more
equations expressing the constraints on the coordinates.
So, requiring a distance $d$ between two points
$A=(A_x,A_y)$ and $B=(B_x,B_y)$ in 2-space would be expressed by
$$
  (A_x-B_x)^2 + (A_y-B_y)^2 = d^2
$$
Requiring that the two points are coincident would entail two
equations:
$$
  \begin{array}{rcl}
    A_x &=& B_x\\
    A_y &=& B_y
  \end{array}
$$
Accordingly, a GCS can be viewed simply as a system of equations: A
solution of the GCS, if one exists, is a valuation of the variables,
the coordinates,
that satisfies all equations.  Viewed in this foundational way,
solving a GCS boils down to formulating a system of equations in the
coordinates of the geometric entities and solving the system by any
means appropriate.  The equations are almost always algebraic.

\subsection{Constraint Graph, Deficit, and Generic Solvability}

The approach of treating a GCS as an (unstructured) system of equations is inefficient and almost always unnecessary.
Given a GCS, we analyze the structure of the
corresponding equation system, and seek to identify smaller subsystems
that can be solved independently and that admit especially efficient
solution algorithms.  Triangle decomposable systems, the main subject of this
\paper, do this by analyzing a {\em constraint graph} that mirrors the equation
structure.  The constraint graph is recursively broken down into
subgraphs that correspond to independently solvable subsystems.  Once
subsystems have been solved, their solutions are combined and expose,
in the aggregate, additional subsystems that now can be solved
separately.  Given the GCS problem $P=(U,F)$, with the set of
geometric objects $U$ and constraints $F$ upon them, we define the
constraint graph $G(P)$ as follows:

\begin{definition}\label{def:consGraph}
Given the GCS $P=(U,F)$, its {\em constraint graph} $G=(V,E)$ is a
labeled undirected graph whose vertices are the geometric objects in
$U$, each labeled with its degrees of freedom.  There is an edge
$(u,v)$ in $E$ if there is a constraint between the geometric objects
$u'$ and $v'$, corresponding to $u$ and $v$ respectively.  The edge is
labeled by the number of independent equations corresponding to the
constraint between $u'$ and $v'$.
\end{definition}

\begin{figure}
  \begin{center}
  \mbox{\subfigure[]{\includegraphics[scale=0.8]{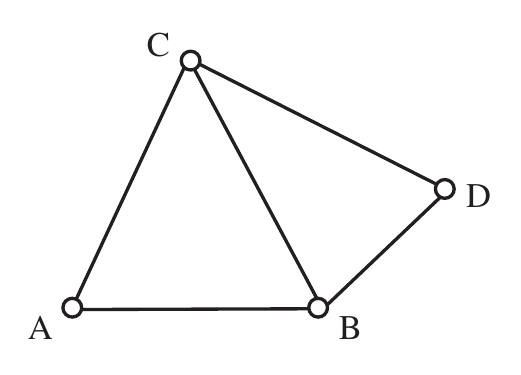}}
      \qquad
      \subfigure[]{\includegraphics[scale=0.8]{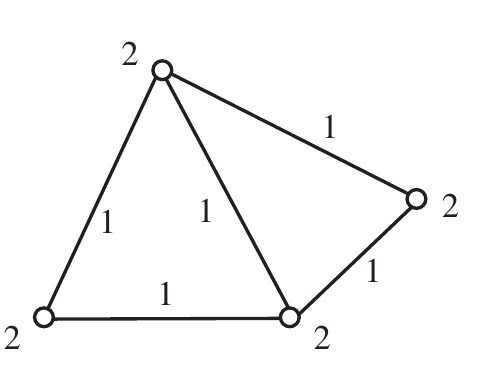}}
       }
  \end{center}
  \caption{Constraint problem and associated constraint graph.}
  \label{fig:truss}
\end{figure}

\begin{example}\label{ex:truss}
Consider the constraint problem of Figure \ref{fig:truss} that
comprises four points in the plane, labeled $A$ through $D$.  A line
between two points indicates a distance constraint on the points.
Thus there are 5 distance constraints, shown left.  The corresponding
constraint graph is shown on the right.
\end{example}

Since a GCS naturally corresponds to a system of equations,
we expect that the
number of independent equations should equal the number of
variables.  The number of independent variables equals the sum of
degrees of freedom, i.e., the sum of vertex labels of the constraint
graph.  Moreover, the number of independent equations equals the sum
of the edge labels, in most situations.  Thus, we investigate ho the structure
of the constraint graph reflects the structure
of the equation system that corresponds to the GCS.

As stated, the GCS of Example~\ref{ex:truss} does not prescribe where
on the plane to position and orient the points after solving the GCS.
Thus, while this GCS is well-constrained, the sum of vertex labels
equals the sum of edge labels plus 3.  Here, the {\em deficit} of 3
corresponds to 3 missing equations that fix where, in the plane, to place the solution.
If necessary, the remaining degrees of freedom can be determined by placing the points
with respect to a global coordinate system, for example by adding three
equations that place $A$ at the origin and $B$ on the (positive) $x$-axis.

So, if the position and orientation of the solution is undetermined, we are assured
that a constraint graph where $\sum l(v) - \sum l(e) < 3$ in the plane corresponds to an equation system in which at least one equation is not independent.
Consequently, we conclude

\begin{definition}
	\label{def:genovercon} 
	A GCS problem $P$ is generically over-constrained if there is an induced subgraph
	of the associated constraint graph, such that for the induced subgraph the following holds:
	$\sum l(v) - \sum l(e) < \kappa$ and none
	of the constraints fixes the geometric structure with respect to the global
	coordinate system, where $\kappa=3$ in the plane and $\kappa=6$ in 3-space.
\end{definition}

Extrapolating this line of reasoning, we might be led to the conclusion that we can use this structural graph property to define 

\begin{definition}\label{def:genundercon} 
	A GCS problem $P$ in the Euclidean plane/space is generically
	under-constrained if it is not overconstrained and $\sum l(v) - \sum l(e) > \kappa$.
\end{definition}

\begin{definition}\label{def:genwellcon} {\bf (Provisional)}\newline
A GCS problem $P$ in the Euclidean plane/space is generically
well-constrained if: (i) it is not over-constrained and (ii) the sum of vertex labels of the associated
constraint graph equals the sum of the edge labels plus the deficit $\kappa$.  
\end{definition}

Note, however, for a graph to be well-constrained Definition~\ref{def:genwellcon} is not sufficient and will be refined in Section \ref{sec:gensolChurch}.  The problem with Definition~\ref{def:genwellcon} is best illustrated by an example.
The following example exhibits a constraint graph for which the sum of the
labels of the vertices equals the sum of the labels of the edges plus $\kappa$ but it 
contains an overconstrained subgraph and therefore the GCS cannot be well-constrained.

\begin{figure}
  \begin{center}
    \includegraphics[scale=0.7]{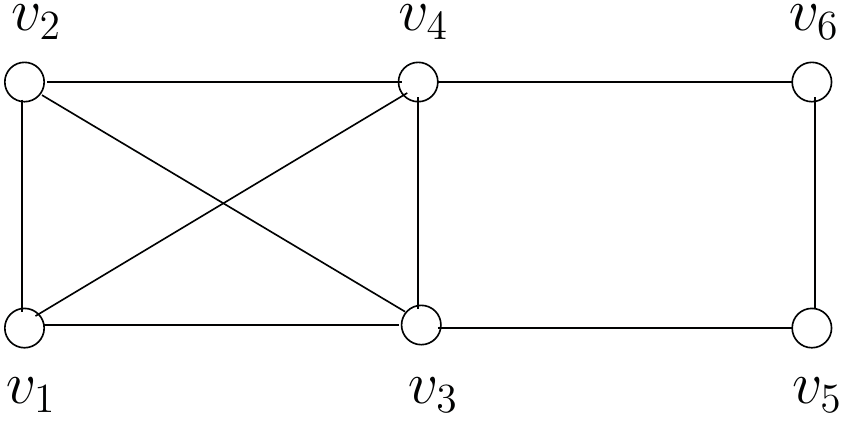}
  \end{center}
  \caption{A GCS with an overconstrained and an underconstrained 
          subgraph.}
  \label{fig:provisional}
\end{figure}

\noindent

\begin{example}\label{ex:provisional}
Consider the constraint graph of Figure~\ref{fig:provisional}.  
For specificity, let the graph vertices be points and the edges distances.  
The deficit is 3, so it seems that a problem with this constraint graph is well-constrained.  
Now the subgraph induced by vertices $v_1,...,v_4$ is overconstrained; drop any of the subgraph edges and you obtain a well-constrained subproblem with a deficit 3.  
On the other hand, the subgraph induced by the vertices $v_3...v_6$ has a deficit of 4 and is in fact underconstrained; vertices $v_5$ and $v_6$ cannot be placed. 
There is an extra constraint in the first subproblem that is not {\em available} for the second subproblem.  
\end{example}

Even in 2D, dof counting fails to account for well-constrained graphs.
We will further study the concept of well-constrained graphs in Section~\ref{sec:gensolChurch}.
Analogously, a GCS in 3-space with deficit 6 that is not overconstrained likewise does not always have a well-constrained graph.

Note that the definitions and properties explained assume in particular that the GCS
solution does not have certain symmetries.  For instance, consider
placing concentrically two circles of given radii.  The constraint
graph has two vertices, labeled 2 each, and one edge labeled 2,
because concentricity implies that the two centers coincide.  Here the
deficit $\sum l(v) - \sum l(e)$ is 2, less than $\kappa=3$.  The
system appears to be over-constrained, but the constraint system is
well-constrained.  The lower deficit reflects the rotational symmetry
of the solution.

In the following, we exclude GCS that fix the structure with respect
to the coordinate system, as well as GCS that exhibit symmetries that
reduce $\kappa$.

\subsection{Instance Solvability}

Consider constructing a triangle $\triangle(A,B,C)$ with the three
sides $a, b, c$ through each of the vertex pairs, in the Euclidean
plane.  The three points and three lines together have 12 degrees of
freedom.  Each vertex is incident to two lines, so there are six
incidence constraints, each contributing one equation.  We add three
angle constraints, one for each line pair.  The resulting constraint
graph has a deficit of three and thus appears to be generically well-constrained.
However, the problem is not well-constrained: if the three angle
constraints add up to $\pi$, then the GCS is actually
underconstrained.  If they add up to $\pi/4$, say, the GCS has no
solution.

The problem arises from the interdependence of the three
angle constraints.  Thus generic solvability does not guarantee that
problem instances are solvable.  In particular, the constraint graph
does not record specific dependencies among the equations.  Such dependencies often arise from geometry theorems.

\begin{definition}
A GCS is (instance) well-constrained if the associated (algebraic)
equation system has one or more discrete real solutions.  It is
(instance) over-constrained if it has no solution, and is (instance)
under-constrained if it has a continuum of solutions.
\end{definition}

The definition of generic solvability is fair in the sense that
dependencies among some of the constraints, and their equational form,
can arise from geometry theorems.  The example above is based on a
simple theorem, but more complex theorems can arise and may be as hard
to detect as solving the equations in the first place.

\subsection{Root Identification and Valid Parameter Ranges}

\begin{figure}
\begin{center}
\mbox{\subfigure[]{\includegraphics[scale=0.7]{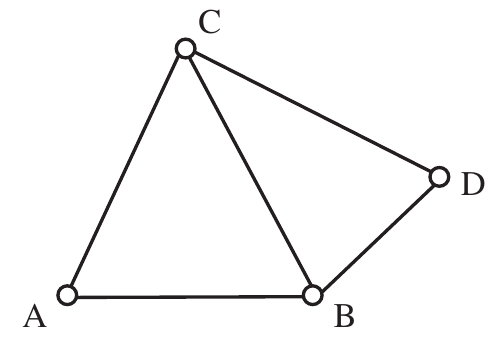}}
      \qquad
      \subfigure[]{\includegraphics[scale=0.7]{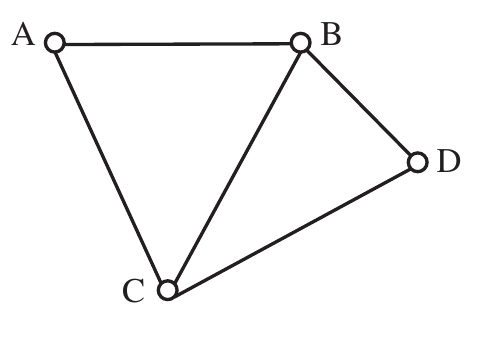}}
      \qquad
      \subfigure[]{\includegraphics[scale=0.7]{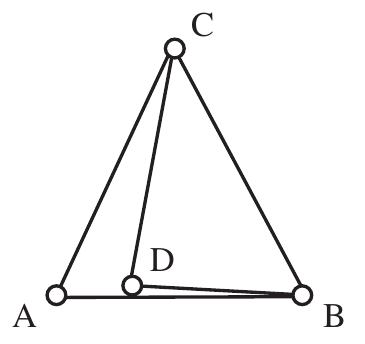}}
     }
\end{center}
\caption{Different solutions of the constraint problem of Example
         {\protect\ref{ex:truss}}.}
\label{fig:truss2soln}
\end{figure}

Consider Example~\ref{ex:truss}. With the distance constraints as
drawn, the GCS has multiple solutions, some shown in Figure
\ref{fig:truss2soln}.
From the equational perspective none is distinguished.  From an
application perspective usually one is intended.  Since the number of
distinct solutions can grow exponentially with the number of
constrained geometric objects, it is not reasonable to determine all
solutions and let the application choose.  This application-specific
problem is the {\em root identification problem}.  Some authors use the term {\em chirality}.

Solvability generalizes to the determination of
{\em valid parameter ranges}: It is clear that the distances between
the points $A,B,C$, and the distances between the points $B,C,D$ must
satisfy the triangle inequality for there to be a (nondegenerate)
solution.  Considering the metric constraints (distance, angle) as
coordinates of points in a configuration space, what is the manifold of
points that are associated with solvable GCS instances?  Is the
manifold connected or are there solutions that cannot be reached from
every starting configuration?

\subsection{Variational and Serializable Constraint Problems}
\label{sec:intro:serial}

Some GCS have equation systems that are naturally triangular.  That
is, intuitively, the geometric objects can be ordered such that they
can be placed one-by-one.  Such GCS are {\em serializable or
  sequential constraint problems}.  GCS that are not serializable have been
called {\em variational constraint problems} in some of the application-oriented literature.  The examples discussed
so far are all sequential.  The following example shows a variational
problem.

\begin{figure}
\centerline{
  \includegraphics[scale=0.7]{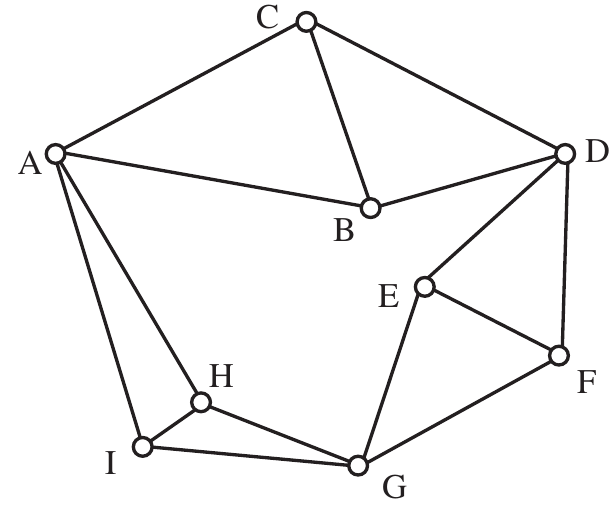}
}
\caption{A variational constraint problem.}
\label{fig:3trusses}
\end{figure}

\begin{example}\label{ex:3trusses}
Place 9 points in the plane subject to the 15 distance constraints
indicated by the lines, in Figure \ref{fig:3trusses}.  The following
three groups of 4 points each can be placed with respect to each
other: $(A,B,C,D)$, $(D,E,F,G)$, and $(G,H,I,A)$.  Each of these subproblems
is sequential in nature.  However, the overall placement problem of
all points is not serializable; it is variational.
\end{example}

Serializable constraint problems are of limited expressiveness.  Due
to their potential efficiency, however, they are used to great effect
in some dynamic geometry packages such as
Cinderella~\cite{bib:cinderella2} and GeoGebra~\cite{bib:geogebra07}.

\subsection{Triangle-Decomposing Solvers}

The overall strategy of triangle-decomposing solvers is to construct the
constraint graph and, analyzing the graph, to recursively isolate
solvable subproblems.  Then, the solved subproblems are recombined.
The process is recursive.

For planar constraint systems,
{\em triangle-decomposability} is understood to be a recursive property in which
the base case consists of two vertices (2 dof each) and one edge
between them eliminating one of the four dof's.  Larger structures are
then solved by isolating three solved structures that pairwise share a single
geometric element and thus can be combined by placing the solved structures relative to each other.
When this decomposition is done top-down, the solved structures are triconnected; \cite{bib:owen91}.
When the decomposition is done bottom-up, triples of solved components are sought
that pairwise share a geometric primitive; \cite{bib:bouma95}.
The act of combining such triples has been called a
{\em cluster merge}.

For the problem of Example
\ref{ex:3trusses}, illustrated in Figure \ref{fig:3trusses},  the graph is first decomposed into the three
subproblems discussed.  Next, each of these subproblems is further
broken down.  For the subproblem $A,B,C,D$, for example, this might be
done by placing, with respect to each other, $A$ and $B$, $B$ and $C$,
and $C$ and $A$.  Those three subsystems are then combined into the
triangle $A, B, C$.  This merging step combines three subsystems that
pairwise share a point, placing the three shared geometric
elements in one geometric construct.  The triangle can then be merged
with the two subsystems $C$ and $D$, and $B$ and $D$.
The result so obtained is the solved subsystem with vertices
$A, B, C, D$.
Similarly, we solve subproblems $D,E,F,G$ and
$G,H,I,A$, each in two merge stpdf.
Finally, the entire problem is solved by merging the three subsystem.

The recursion solving the problem of Figure \ref{fig:3trusses}
can be thought of either as a top-down decomposition, or as a
bottom-up reconstruction.

Note that the top-level merge step places the shared objects, here the
points $A$, $D$ and $G$.  The distances between these points are not
given but can be obtained from the three subproblem solutions.  We can
think of this process as a plan that is formulated based on the
constraint graph analysis.  Two questions arise:
\begin{enumerate}
\item
  Since the plan so formulated is not unique, do different
  decompositions arrive at different solutions?  This question is
  settled by investigating the nature of the recursive decomposition
  and whether it satisfies a Church-Rosser property.
\item
  What specific subsystems of equations must be solved and how?  This
  question is approached by analyzing the basic subgraph
  configurations that can occur, that are allowed by the geometric
  constraint solver, both when decomposing and when recombining.
\end{enumerate}
Concerning the second question, in our example two key operations are
needed: placing two points at a prescribed distance, and placing a
third point at a required distance from two fixed points.  There is
also a third operation that places a rigid geometric configuration $X$
such that two points of $X$ match two given points, using translation
and rotation.  This operation is used when recombining the three
subproblems.

Carrying out these operations entails solving equation systems of a
fixed structure.  Often, these equation systems are small and can be
solved very efficiently.  The simplest cases restrict the geometric repertoire
to points, lines, and circles of fixed radius.  Triangle decomposable systems
then require at most solving univariate quadratic polynomials.

More complicated primitives can be added.  They include circles of
a-priori unknown radius, e.g., \cite{bib:kavasseri96};
conic sections, e.g., \cite{bib:fudos96};
and B{\'e}zier curves, e.g., \cite{bib:HofPe95}.
Additions impact both the constraint graph analysis as well as the richness
of algebraic equation systems that have to be solved.
The impact on the graph analysis can be limited to adding additional patterns to cluster merging.
So extended, triangle decomposition becomes a more general analysis that can be called
{\em tree decomposition}.  The decomposition remains a tree, but the tree may have different numbers of subtrees associated with
a growing repertoire of subgraph patterns.
The number of possible patterns is infinite.  Therefore, this approach becomes self-limiting.
More than that, rigidity no longer has a simple characterization.
The contributions to the equation solving algorithms is also considerable.
Section \ref{sec:varraded} catalogues what is known about the equation systems required to add just circles of unknown radius.

Because of these rigors, more general techniques have to be considered.
For the extended graph analysis, the DR-type algorithms solve the problem
by dynamically identifying generically solvable subgraphs; \cite{hoffman2001decomposition1,hoffman2001decomposition2}.
Likewise, the growth of irreducible algebraic equation systems increasingly
motivates searching general equation solvers, including
numerical solving techniques, such as Newton iteration,
homotopy continuation, and other procedural techniques,
for instance \cite{bib:SherPa93}.

\subsection{Scope and Organization}

We begin with the graph analysis of triangle-decomposable constraint
systems in 2D.  These systems play an important role in linkage
analysis and graph rigidity,
\cite{bib:sitharam14,bib:sitharamUa,bib:sitharamUb}.
But even in the Euclidean plane, applications of geometric constraint
systems argue for more expressive systems.  We can increase
expressiveness by enlarging the repertoire of geometric objects, as
well as by admitting more complex cluster merging operations.

In the plane, an extended repertoire includes foremost variable-radius
circles; that is, circles whose radius is not given explicitly but
must be inferred based on the constraints.  More advanced objects,
such as conics and certain parametric curves, can also be considered.
Those additional geometries impact the subsystems of equations that
must be solved.

More complex cluster merging operations affect both the constraint
graph structures encountered as well as the equation systems that must
be solved to position the geometric structures accordingly.  We will
discuss some examples.

After discussing constraint graphs and generic over-, well-, and
under-constrained graphs, we consider the equations that must be
solved so as to obtain specific solutions.  Here, we explain the
structure of those systems as well as some tools to transform the
equation systems into simpler ones.  Questions of root identification,
valid parameter ranges, and order type of solutions are to be
discussed.  Much is known about these topics in the Euclidean plane.
For Euclidean 3-space much less is known.  There the equation systems are in
general much harder, and the number of cases that should be considered
is larger.  Even simple sequential problems can require daunting equation systems.
A case in point is to find a line in 3-space that is at prescribed distances from four given points, discussed in Section \ref{sec:C3dSeqLine}.


\section{Geometric Constraint Systems}

\begin{definition}
A {\em geometric constraint system} (GCS) is a pair $(U,F)$ consisting
of a finite set $U$ of geometric elements in an ambient space and a
finite set $F$ of constraints upon them.
\end{definition}
The ambient space typically is $n$-dimensional Euclidean
space. The majority of applications require $n=2$ or $n=3$; e.g.,
\cite{bib:owen91,bib:bouma95,bib:hoffmann00}.  Spherical geometry may
also be considered, for instance in nautical applications.

The imposed constraints typically are binary relations.  We do not
consider higher-order constraints, such as "$C$ be the midpoint
between points $A$ and $B$."  Note, however, that such constraints can
often be expressed by several binary constraints.
This can be done in a variety of ways, with or without variable-radius circles.

\begin{definition}
A {\em solution} of a geometric constraint system $(U,F)$ is an
assignment of coordinates instantiating the elements of $U$ such that
the constraints $F$ are all satisfied.
\end{definition}
There may be several solutions \cite{bib:fudos96b}.
Moreover, solutions may or may not be required to be in prescribed position and
orientation, in a global coordinate system.

As defined, a GCS is a static problem in that solutions fix the
geometric elements with respect to each other.  The
\emph{dynamic geometry} problem asks to maintain constraints as some
elements move with respect to each other.  We consider only static
constraint problems and their solvers.

By \emph{geometric coverage} we understand the diversity of geometric
elements admissible in $U$.  Points, lines and circles of given radius
are adequate for many applications in Euclidean 2-space
\cite{bib:owen91}.  For GCS in Euclidean 3-space, an analogous geometric
coverage could be points, lines, planes, as well as spheres and
cylinders of fixed radii.  Here, the number of solutions even of simple GCS
can be very large \cite{bib:gao04b}.


\section{Constraint Graph}\label{sec:CGraph}

The constraint graph of Definition~\ref{def:consGraph} is an abstract
representation of the equation system equivalent to the geometric
constraint problem.  The analysis of the graph yields a set of
operations used to solve the equations.  Ideally, those operations
are simple, for instance univariate polynomials of degree 2 \cite{bib:bouma95}.
To start the graph analysis, we find
{\em minimal constraint problems}.
That is, constraint problems with a minimal number of geometric objects whose solution defines a local coordinate frame.
Such problems depend on ambient space.  Consider the following:

\begin{definition}\label{def:minimalCG}
Given a constraint problem in the Euclidean plane, consisting of two points $A$ and $B$ and a nonzero distance constraint between them.
Such a problem is {\em minimal}.  The associated constraint graph $G=(\{A,B\}, \{(A,B)\})$ is a {\em minimal constraint graph}.
\end{definition}

This minimal constraint problem establishes a coordinate system of the Euclidean plane in which $A$ is the origin and the oriented line $\overrightarrow{AB}$ is the x-axis.
This is not the only minimal constraint problem in the plane.  Table \ref{tab:minimalGCS2d} shows the minimal problems involving the basic geometric objects with 2 dof.
Note that fixed-radius circles can be used in lieu of points as long a the centers and points are not coincident.
Two parallel lines, at prescribed distance in the plane, are not considered minimal because they do not establish a coordinate frame.

\begin{table}
\begin{center}
\begin{tabular}{| l | l | l |}\hline
{\bf Geoms} & {\bf Constraint} & {\bf Notes} \\ \hline
Points $A,B$ & $d(A,B)$ & distance not zero \\ \hline
Point $A$, line $m$ & $d(m,A)$ & zero distance allowed \\ \hline
Lines $m,n$ & $a(m,n)$ & lines not parallel \\ \hline
\end{tabular}
\end{center}
\caption{Minimal GCS in the Euclidean plane; $d(...)$ denotes distance, $a(...)$ denotes angle.}
\label{tab:minimalGCS2d}
\end{table}

Table \ref{tab:minimalGCS3d} shows the main cases for Euclidean 3-space.  In the case of two lines that are skew, a third line $L_3$ is constructed that connects the two points of closest approach.  Here, $L_1$ and $L_3$ define a plane that is oriented by $L_2$.
If the lines $L_1$ and $L_2$ intersect, they lie on a common plane that is coordinatized by the two lines and is oriented by defining the third coordinate direction using a right-hand rule.
If the two lines are parallel, they define a common plane but fail to coordinatize it.

\begin{table}
\begin{center}
\begin{tabular}{| l | l | l |}\hline
{\bf Geoms} & {\bf Constraint} & {\bf Notes} \\ \hline
Points $p_1 , p_2 , p_3$ & $d(p_i , p_k)$ & no zero distance \\ \hline
Point $p$, line $L$ &	$d(p,L)$ & distance not zero \\ \hline
Lines $L_1 , L_2$ & $d(L_1,L_2), a(L_1,L_2)$ & lines not parallel \\ \hline
Planes $E_1 , E_2 , E_3$ & $a(P_i , P_k)$ & no parallel planes \\ \hline
\end{tabular}
\end{center}
\caption{Minimal GCS in Euclidean 3-space; $d(...)$ denotes distance, $a(...)$ denotes angle.}
\label{tab:minimalGCS3d}
\end{table}

%
%


\subsection{Geometric Elements and Degrees of Freedom}
\label{sec:geoDOF}

A geometric element has a characteristic number of degrees of freedom
that depends on the type of geometry and the dimensionality of ambient
space.

\begin{definition}
Degrees of freedom (dof) are the number of independent,
one-dimensional variables by which a geometric object can be
instantiated and positioned.
\end{definition}

Elementary objects are geometric objects that have a certain number of
dof and cannot be decomposed further into more elementary objects.
Compound objects can be characterized by a GCS, and consist of one or
several elementary objects placed in relation to each other according
to a GCS solution instance.  A complex object is rigid if its shape
cannot change, or, equivalently, if the relative position and
orientation of the elementary objects that comprise the complex object
cannot change.

The degrees of freedom for elementary geometric objects and for rigid
objects in two and three dimensions are summarized in
Table~\ref{tab:dof}.  Singularities of coordinatization can trigger
robustness issues in GCS. Therefore, the representation of elementary
geometric objects should be uniform, without singularities.  For
example, if we represent lines by the familiar  $y=mx+b$ formula, lines parallel to the $y$-axis cannot be so represented.
This problem can be avoided if we represent a line by
its distance from the origin and
the direction of the normal vector of the line.

\begin{table}[tbp]
  \begin{center}
  {\small
    \begin{tabular}{| l | l | c | c |}\hline
     {\bf Geom} & {\bf Geometric Meaning of DoF} & {\bf 2D} & {\bf 3D} \\ \hline
     Point & Variables representing coordinates & 2 & 3 \\ \hline
     Line  & 2D: distance from origin and direction & & \\
           & 3D: distance from origin, direction in 3D & 2 & 4 \\
           & direction on the plane && \\ \hline
     Plane & Distance from origin, direction in 3D && 3 \\ \hline
     Circle, fixed-radius & Coordinates of center, orientation in 3D & 2 & 5 \\ \hline
     Circle, variable-radius & Coordinates of center, radius, orientation in 3D & 3 & 6\\ \hline
     Rigid body & 2D: 2 displacements, 1 orientation & 3 & 6 \\
                & 3D: 3 displacements, 3 orientation && \\ \hline
     Sphere, fixed- or & 3 displacements or & 3 & 4\\
     variable-radius   & 3 displacements, radius &&\\ \hline
     Ellipse, variable axes & center, axis lengths, axis orientation & 5 & 7 \\ \hline
     Ellipsoid, variable axes & center, axis lengths, axis orientation & & 9 \\ \hline
    \end{tabular}
    }
  \end{center}
  \caption{Degrees of freedom for elementary objects and rigid objects.}
  \label{tab:dof}
\end{table}

\begin{table}
  \begin{center}
  {\small
    \begin{tabular}{| l | l | c | c |}\hline
    \textbf{Type} & \textbf{Constraint} & \textbf{2D}
                                        & \textbf{3D}  \\ \hline

    point-point distance         & \begin{minipage}[t]{6cm}
                                      One equation representing the
                                      distance between two points
                                      $p_1,p_2$ under the metric
                                      $||\ ||$: 
                                      $|| p_1 - p_2|| = d$ with $d>0$.
                                      \end{minipage}
                                    & 1  &1  \\ \hline
    Angle between lines and planes  & \begin{minipage}[t]{6cm}
                                      Angle between two lines (can be
                                      represented by the angle between
                                      the normal vectors). \\
                                      \textbf{Exceptions in 3D:} Prallelism
                                      between lines eliminates 2DoF.
                                      Line-plane orthogonality in
                                      eliminates 2 DoF.
                                      \end{minipage}
                                      & 1 & 1  \\ \hline

    Point on point                  & \begin{minipage}[t]{6cm}
                                      For any metric $||p_1 - p_2|| =0$
                                      is equivalent to $p_1 = p_2$.
                                      \end{minipage}
                                    & 2 & 3  \\ \hline

    Point-line distance             & \begin{minipage}[t]{6cm}
                                      One equation expressing the
                                      point-line distance.
                                      \end{minipage}
                                    &1 & 1  \\ \hline

    Line-line parallel distance     & Parallelism and distance.
                                         & 2  & 3  \\ \hline

    Plane-plane parallel distance   & Parallelism and distance.
                                         & 2  & 3  \\ \hline

    Point on line                        & \begin{minipage}[t]{6cm}
                                           Same as point-line distance
                                           In 3D dimension is reduced.
                                           \end{minipage}
                                         & 1 & 2  \\ \hline

    Line on line                         & \begin{minipage}[t]{6cm}
                                           Same as parallel distance
                                           between lines in 2D. In 3D
                                           an additional DoF is canceled.
                                           \end{minipage}
                                         & 2 & 4  \\ \hline

    Point on plane                       & Same as point-plane distance.
                                         & - & 1  \\ \hline

    Line on plane                        & Plane-line parallelism and
                                           zero  distance.
                                         & - & 2  \\ \hline

    Fixing elementary object             & \begin{minipage}[t]{6cm}
                                           Fixing all or some of the
                                           DoF of an elementary object.
                                           \end{minipage}
                                         & Dof & DoF  \\ \hline
    \end{tabular}
  }
  \end{center}
  \caption{Types of constraints and number of dof eliminated.}
  \label{tab:constraints}
\end{table}

\subsection{Geometric Constraints}
\label{sec:geoDOF_2}

Geometric constraints consume one or more dof and are expressed by an
equal number of independent equations. A basic set of geometric constraints in the plane is
summarized in Table~\ref{tab:constraints}.  Constraints can be
expressed directly.  Alternatively, some can be reduced equivalently
to other constraints.  For example, suppose we stipulate that a circle
of radius $r$ be tangent to a line $L$ in the plane.  Then we can
require equivalently that the center of the circle be at distance $r$
from $L$.  For a comprehensive list of constraints involving planes,
points and lines in 3D see for example
\cite{bib:lee2013,bib:trombettoni2006}.  Finally we should note that
distance 0 between two points takes out two degrees of freedom since
this causes dimension reduction (from 1-dimensional to 0-dimensional).
Similarly, requiring that two lines be parallel at distance $d$ in
3-space eliminates 3 dof, but if $d=0$ the constraint eliminates 4
dof.

\subsection{Compound Geometric Elements}
\label{sec:geoDOF_3}

\begin{figure}
  \begin{center}
  \mbox{\subfigure[]{\includegraphics[scale=1.0]{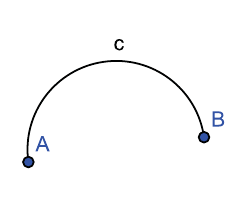}}
    \subfigure[]{\includegraphics[scale=0.7]{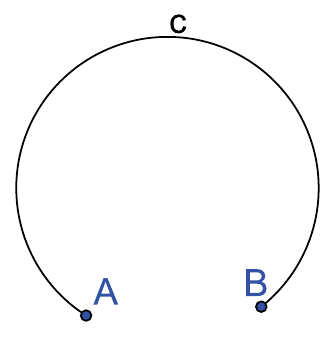}}
    \subfigure[]{\includegraphics[scale=1.0]{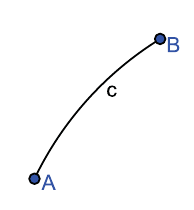}}
    \subfigure[]{\includegraphics[scale=0.4]{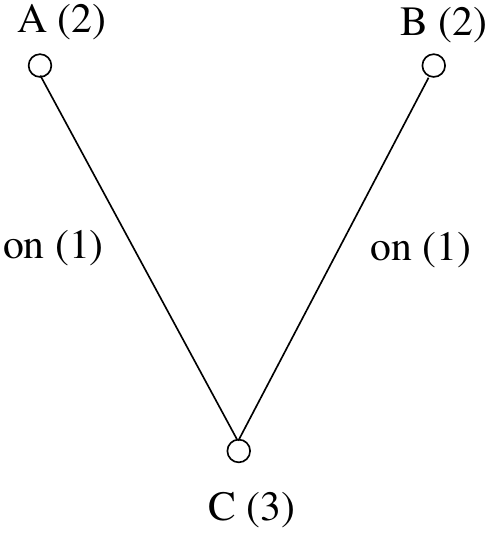}}
  }
  \end{center}
  \caption{Building a complex object using one variable radius circle,
           two end points and two incidence constraints.}
  \label{fig:arc}
\end{figure}

Several geometric elements can be conceptually grouped into compound elements.
A circular arc is an example.
Compound geometric elements are convenient concepts for graphical user interfaces of GCS solvers.
Internally, they are decomposed into geometric elements and implied constraints.
We illustrate this dual view with the example of a circular arc shown in
Figure~\ref{fig:arc} and explain why it has overall 2 intrinsic parameters.

Since the arc does not have a fixed radius, it lies on a variable-radius circle that has 3 dof.
The two end points contribute an additional 4 dof.
The two end points are incident to the circle, reducing the overall dof to 5.
Position and orientation of the arc, in some
coordinate system, requires 3 dof, leaving two intrinsic arc
parameters.  They can be interpreted as 1 dof for the radius of the
arc, and 1 dof for the distance between the end points.

Note that
this constraint problem has 4 solutions in general.  Orienting the end
points, and connecting them with an oriented line, or circle, the arc can be on
one of two sides of the directed line.  Moreover, the two end points
divide the circle into a shorter and a longer arc, in general.  Which
one is chosen accounts for the other two solutions.  Applications
require selecting one of these solutions.  Several conventions can be
followed to determine a unique segment: preservation of the original
configuration, preservation of the direction of the curve from the
first endpoint towards the second endpoint, the shortest segment and
so on.

\subsection{Serializable Graphs}

When a GCS is serializable (Section \ref{sec:intro:serial})
the geometric objects can be ordered in such a way that they can be
placed sequentially one-by-one as function of preceding, already placed
elements.  This idea can be formalized as follows.

\begin{definition}
Let $G=(V,E)$ be a GCS graph and $x_0, x_1, x_2 \ldots$ be elements in
$V$.  We say that $x_i$ depends on $x_k, x_r, \ldots$, written
$x_i>x_k,x_r,\ldots$, if $x_i$ can be placed only after the $x_k,x_r,
\ldots$ have been placed.
\end{definition}
If the constraint graph is serializable, then the pair $(G, >)$ is a
directed acyclic graph (DAG) and admits topological
sorting~\cite{bib:aho74}.
See Example \ref{ex:3trusses}.  More formally,

\begin{definition}
Let $G(V,E)$ be the constraint graph associated with a GCS in Euclidean 2-space.
Without loss of generality assume that $G$ is connected.  We say that
the GCS is serializable if $(G, >)$ describes a sequence
of dependencies such that, under a suitable enumeration of $V$,
\begin{enumerate}
\item
There are elements $x_0$ and $x_1 \in V$ that induce a minimal
constraint graph.
\item
Each subsequent element $x_i,~~~2 \leq i \leq |V|$ depends on elements
$x_j$ and $x_k$ where $j<i$ and $k<i$.
\end{enumerate}
\end{definition}
In general, the enumeration is not unique and depends on the pair
$x_0, x_1 \in V$ that is placed first. However, as  we
will see later, different possible sequences derived from a given
DAG are equivalent in the sense that they lead to the same final
placement for all the objects in $V$ with respect to each other;
see Section~\ref{sec:gensolChurch}.

\begin{figure}
  \begin{center}
  \mbox{\subfigure[]{\includegraphics[scale=0.5]
                    {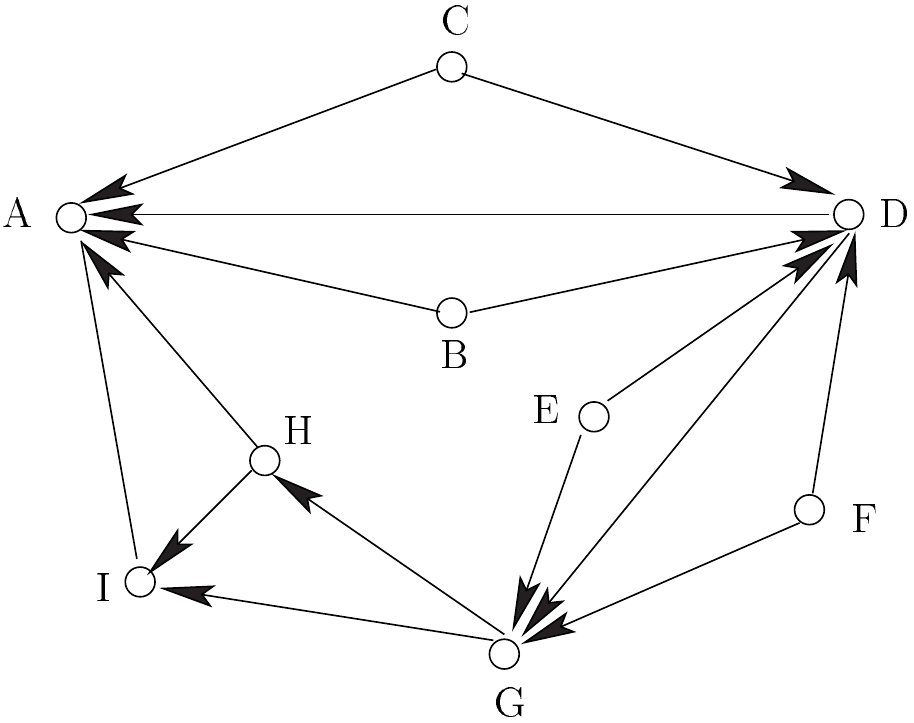}}
        \hspace{1cm}
        \subfigure[]{\includegraphics[scale=0.55]
                    {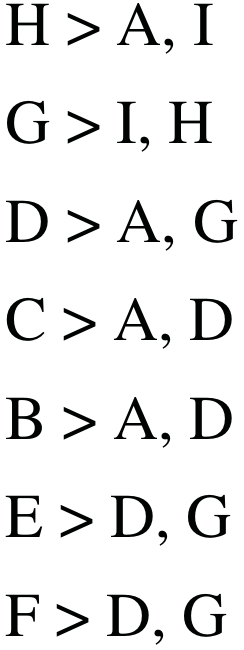}}
  }
  \end{center}

  \caption{a) DAG derived from a serializable graph, with
              $(A, I)$ as starting pair.\hfill\break
           b) A construction sequence respecting the dependencies.}
  \label{fig:serialDAG}
\end{figure}

\begin{example}
Consider the graph in Figure~\ref{fig:serialDAG}(a).  The edges have
been directed to show dependencies of placement.  Choosing $(A,I)$ as
starting pair, a valid dependence relation is obtained.  The list in
Figure~\ref{fig:serialDAG}(b) gives a serial construction based on the
graph $(G,>)$.
\end{example}

\subsection{Variational Graphs}

\begin{definition}
A GCS which is not serializable is called \emph{variational}.
\end{definition}

When the GCS is variational, starting with a minimal GCS and applying the
dependence relationship to the constraint graph $G = (V,E)$ generates
a sequence of dependencies that includes only a subset of elements in
$V$. We call it a \emph{subsequence} and the corresponding subgraph a
{\em cluster}.

Assuming that the variational GCS is solvable, one repeatedly selects a minimal GCS
and applies the dependence relationship, using graph edges not yet used, resulting in a collection of
subsequences.

Intuitively, the situation described means that clusters corresponding
to different sequences must be merged, usually applying translations
and rotations defined by elements shared by subsequences. From an
equational point of view, the existence of different subsequences
reveals that there are several underlying equations that must be
solved simultaneously.

\begin{figure}
  \begin{center}
  \mbox{\subfigure[]{\includegraphics[scale=0.5]
                    {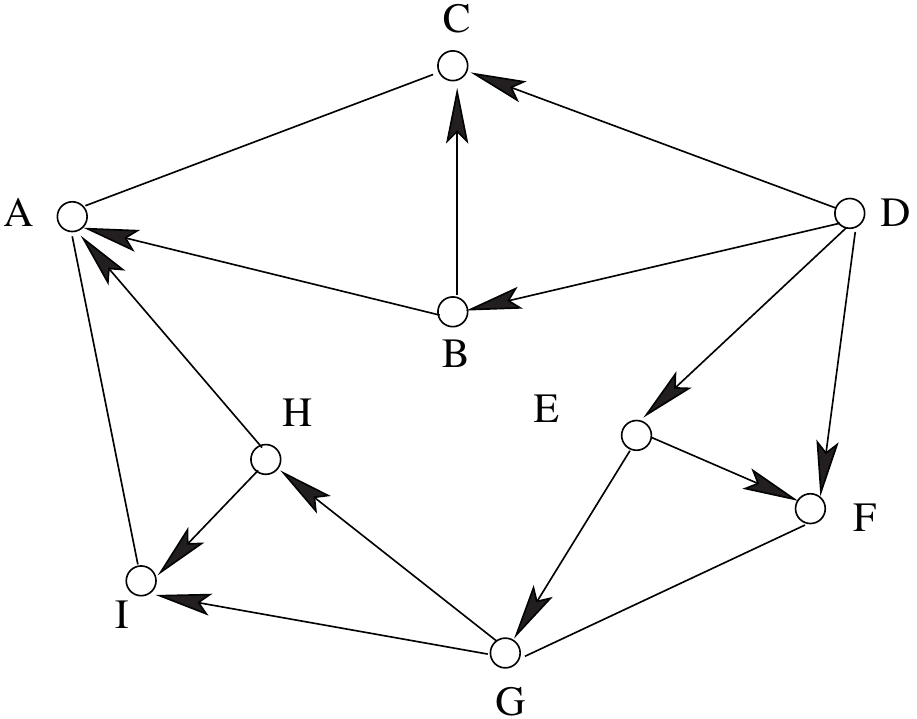}}
        \qquad
        \subfigure[]{\includegraphics[scale=0.45]
                    {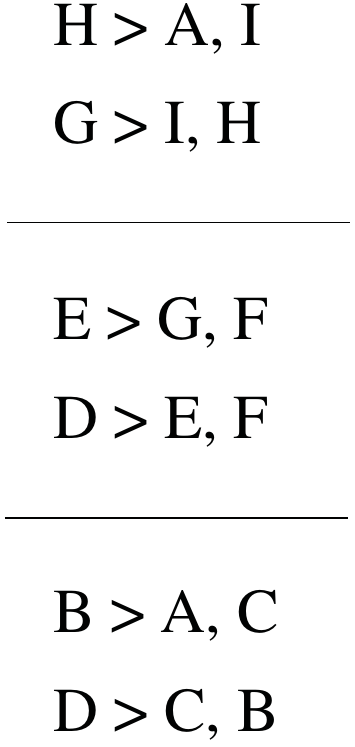}}
  }
  \end{center}

  \caption{a) DAG derived from the variational graph in
              Figure~\ref{fig:3trusses} with starting pairs $(A,I)$,
              $(G,F)$ and $(C,B)$.\hfill\break
           b) Three different subsequences of construction dependencies
              that can be identified in the DAG.}
  \label{fig:variationalDAG}
\end{figure}

\begin{example}
Figure~\ref{fig:variationalDAG}a shows a variational DAG corresponding
to the variational constraint problem in Figure~\ref{fig:3trusses}.
We choose three starting pairs $(A,I)$, $(G,F)$ and $(C,B)$.  Each allows us to
build a DAG from some of the graph vertices and edges.  They are listed in
Figure~\ref{fig:variationalDAG}(b). Notice that each subsequence identifies a
serializable subgraph.
\end{example}

\subsection{Triangle Decomposability}\label{sec:treeDecomp}

\begin{figure}
  \begin{center}
  \mbox{
      \subfigure[]{\includegraphics[scale=0.4]{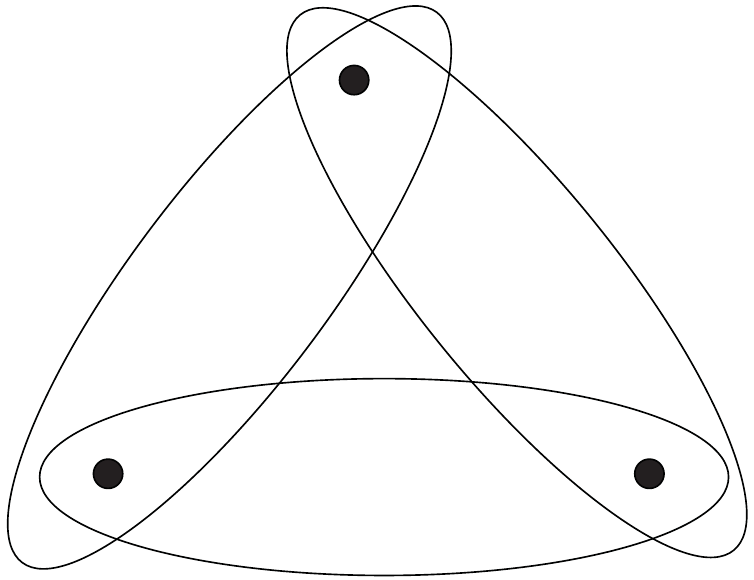}}
      \hspace{2cm}
      \subfigure[]{\includegraphics[scale=0.4]{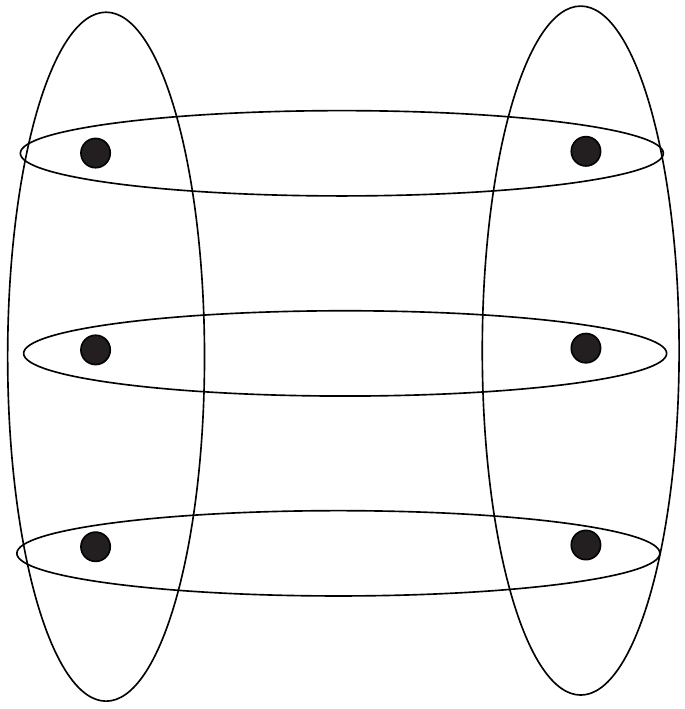}}
  }
  \end{center}
  \caption{a) Triangle decomposition step.
           b) More complex tree decomposition step.
           Split subgraphs are shown as ovals, shared vertices as dots.}
  \label{fig:C23split}
\end{figure}

The strategy of triangle-decomposing solvers, as sketched in
Section~\ref{sec:intro}, is based on decomposing the constraint graph
recursively. Decomposition splits a (sub)graph into
3 (sub)subgraphs that share one vertex pairwise.
This is called a {\em triangle decomposition step}.
More complex splitting configurations can be considered and called \emph{tree decomposition}.
Figure \ref{fig:C23split} shows a triangle and a more complex tree decomposition step.

From a practical point of view, in the case of GCS in Euclidean
2-space, triangle-decomposable constraint graphs suffice for solving
many problems that arise in applications.

We define a \emph{triangle decomposition step} as
follows.

\begin{definition}\label{def:treeDecomp}
Let $G = (V, E)$ be a graph. We say that three subgraphs of $G$,
$G_i(V_i,E_i), 1 \leq i \leq 3$,
define a triangle decomposition step of $G$ if
\begin{enumerate}
\item
 $V_1 \cup V_2 \cup V_3 = V$,
 $E_1 \cup E_2 \cup E_3 = E$ and
 $E_i \cap E_j = \emptyset, i\neq j$,
and

\item
There are three vertices, say $u, v, w \in V$, such that
$V_1 \cap V_2 = \{u\}$,
$V_2 \cap V_3 = \{v\}$ and
$V_3 \cap V_1 = \{w\}$.
\end{enumerate}
\label{def:treedecompA}
\end{definition}

\begin{figure}
  \begin{center}
  \includegraphics[scale=0.5]{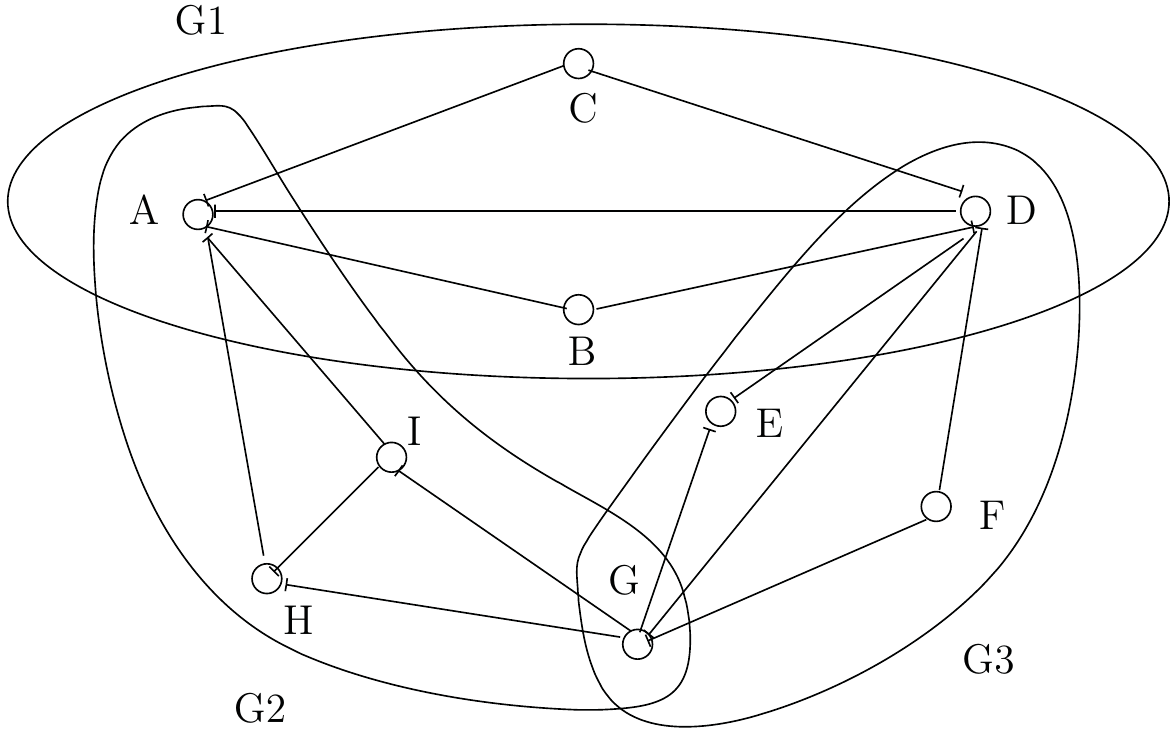}
  \end{center}
  \caption{A triangle-decomposition step for the graph shown
           in Figure~\ref{fig:variationalDAG}(a).}
  \label{fig:decstep}
\end{figure}

\begin{example}
Consider the graph $G$ in Figure~\ref{fig:variationalDAG}(a). As
shown in Figure~\ref{fig:decstep}, the subgraphs $G_1, G_2$ and $G_3$
define a triangle decomposition step of $G$. Vertices pairwise shared by
the subgraphs are $A, D$ and $G$.
\end{example}

\begin{definition}
We say that a ternary tree $T$ is a
\emph{triangle decomposition} for the graph $G$ if
\begin{enumerate}
  \item The root of $T$ is the graph $G$.
  \item Each node of $T$ is a subgraph $G' \subset G$ which is either
    the root of a ternary tree generated by a triangle decomposition step
    of $G'$ or a leaf node with a minimal associated subgraph.
\end{enumerate}
\end{definition}

\begin{definition}
A graph for which there is a triangle decomposition is called
\emph{triangle-decomposable}.
\label{def:treedecompB}
\end{definition}

In general, a triangle decomposition of a graph is not unique.
However, if the graph is triangle decomposable by one sequence of decomposition stpdf,
then any legal sequence will decompose the graph \cite{bib:fudos96b}.

\begin{figure}
  \begin{center}
    \mbox{\subfigure[]
         {\includegraphics[scale=0.5]{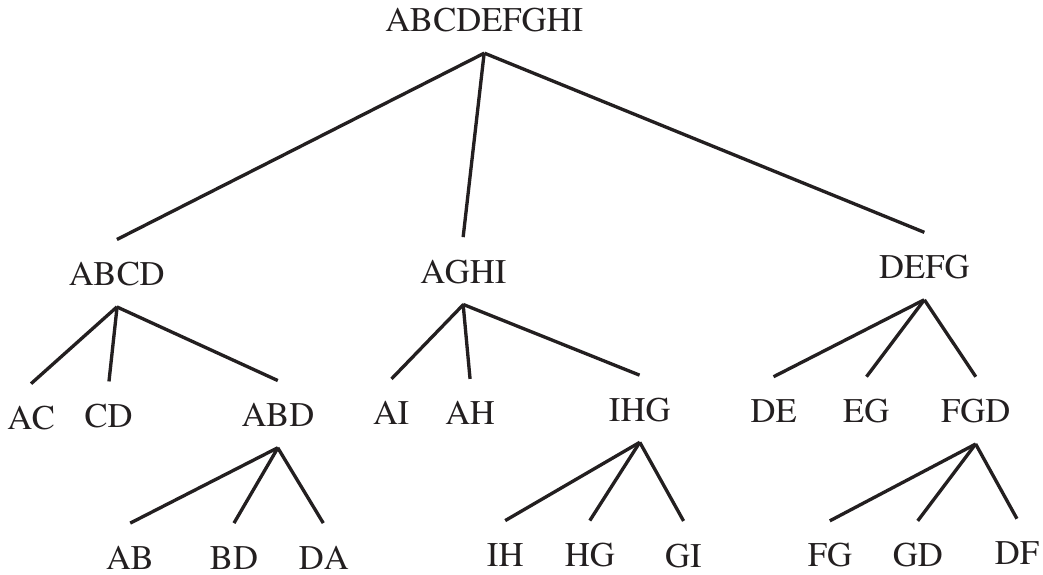}}
         \subfigure[]
         {\includegraphics[scale=0.5]{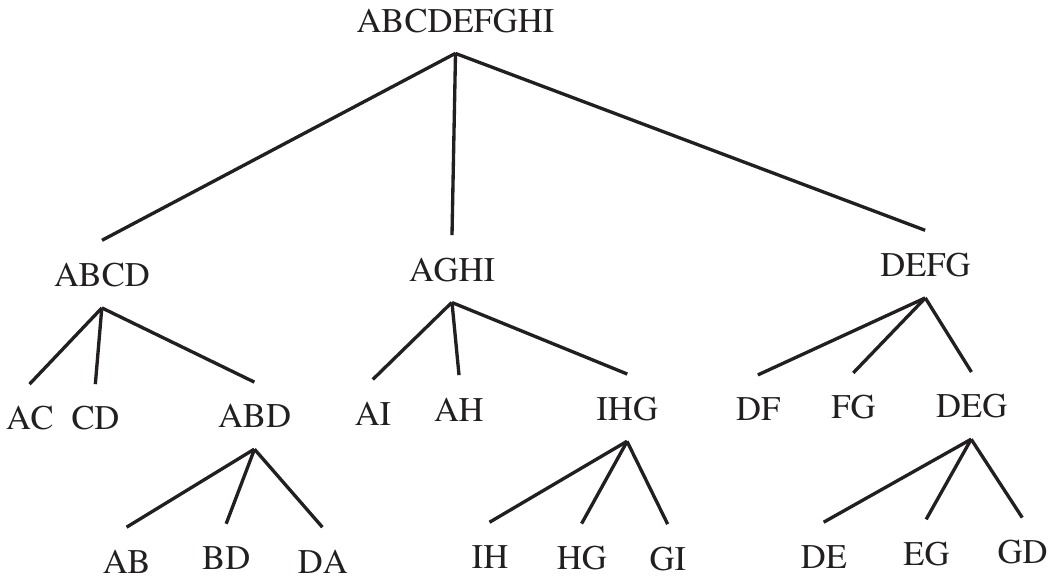}}
    }
  \end{center}

  \caption{Two different triangle decompositions for the graph
           shown in Figure~\ref{fig:variationalDAG}.}
  \label{fig:treedec}
\end{figure}

\begin{example}
Consider the graph and the triangle decomposition step shown in
Figure~\ref{fig:decstep}. Now recursively apply decomposition stpdf to
each of the subgraphs $G_1, G_2, G_3$ until reaching minimal
subgraphs.  Figure~\ref{fig:treedec} shows two triangle decompositions for
the graph considered that differ in the subtree rooted at node
$DEFG$. Notice, however, that the set of terminal nodes is the same
in both triangle decompositions.
\end{example}

Assume that the three subgraphs $G_1, G_2, G_3$ are (graphs of)
solvable GCS subproblems.  Let $u$ and $v$ be the shared vertices of
$G_2$ and there is no constraint between them.  Then $u$ and $v$
constitute a virtual minimal GCS where the constraint relating the two
elements is not given but can be deduced from the solution of $G_2$. A
similar statement can be made about $G_1$ and $G_3$ and their shared
vertices.

The triangle pattern is not the only decomposition construct
\cite{bib:bouma95}.  Others include the pattern shown in Figure
\ref{fig:C23split}b.  Intuitively, a decomposition pattern represents
an equation system that must be solved simultaneously.  Decomposition
patterns are infinite in number; see also 
SECTION 2.2

\subsection{Generic Solvability and the Church-Rosser Property}
\label{sec:gensolChurch}

Recall Definitions \ref{def:treedecompA}, \ref{def:treedecompB}, and
\ref{def:genwellcon}.

\begin{figure}
\begin{center}
  \mbox{\subfigure[]{\includegraphics[scale=0.7]{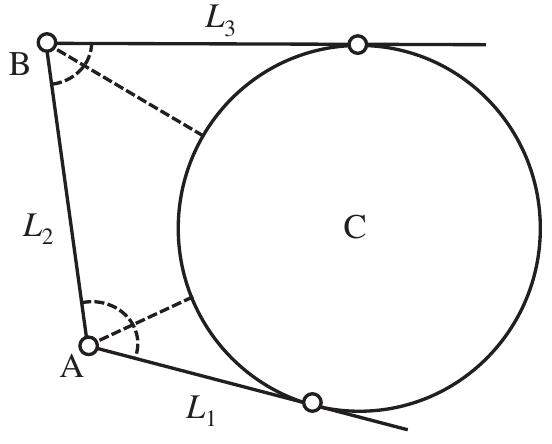}}
        \hspace{1.5cm}
        \subfigure[]{\includegraphics[scale=0.45]{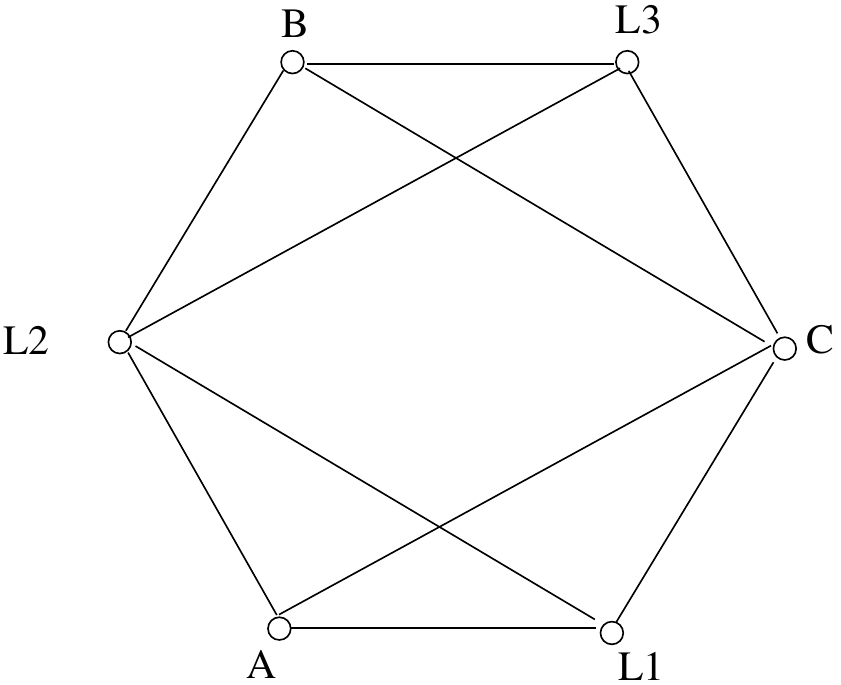}}
  }
\end{center}
  \caption{Well-constrained graph with three lines, two points, and a
  variable-radius circle.
  a) Constraint problem. Dashed lines represent metric constraints.
  b) Constraint graph.  All vertices have 2 dof except C which has
  three. Solid lines represent incidence constraints.
 }
\label{fig:VarCcluster}
\end{figure}

Triangle-decomposable graphs, of GCS in the Euclidean plane, are
generically well-constrained, and can be solved either
bottom-up~\cite{bib:fudos97} or top-down~\cite{bib:joan02}.  This
assertion is based on the shared geometric elements, of each triangle
decomposition step, having 2 dof and being in general position with
respect to each other.  Variable-radius circles have 3 dof and give
rise to a special case illustrated in Figure~\ref{fig:VarCcluster}.

Triangle decomposable graphs are also called
\emph{ternary-decomposable} on account of the topology of the
decomposition tree.  A (recursive) decomposition of a well-constrained
triangle decomposable graph is not unique.  However, it can be shown using
the Church-Rosser property for reduction systems that
if one triangle decomposition sequence fully reduces the graph, then all
such decompositions must succeed,~\cite{bib:fudos96b}.  The advantage
of restricting to triangle-decomposable problems is that a fixed, finite
repertoire of algebraic equation systems suffices to solve this class
of problems.

Triangle decomposable graphs are a subset of the set of well-constrained
constraint graphs of planar GCS.  The entire set has been
characterized by Laman in~\cite{bib:laman70}.
\begin{definition}
Let $G=(V,E)$ be a connected, undirected graph whose vertices
represent points in 2D and edges represent distances between points.
$G$ is a well-constrained constraint graph of a GCS iff, the deficit
of $G$ is 3 and, for every subset $U\subset V$, the induced subgraph
$(U,F)$ has a deficit of no less than 3.
\end{definition}

\begin{figure}
\begin{center}
\mbox{\subfigure[]{\includegraphics[scale=1]{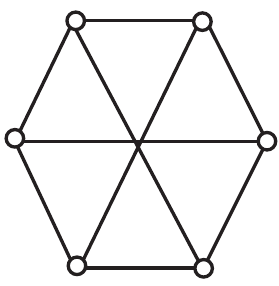}}
      \hspace{1.5cm}
      \subfigure[]{\includegraphics[scale=1]{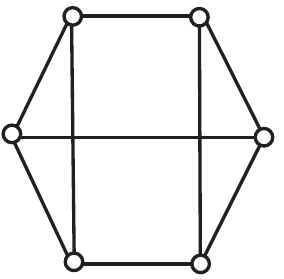}}
}
\end{center}
  \caption{Well-constrained graphs that are not triangle-decomposable.
  a) Graph $K_{3,3}$.
  b) Desargue's graph.  All vertices have 2 dof, all edges consume 1
  dof.
}
\label{fig:Laman}
\end{figure}

Two examples of well-constrained graphs that are not triangle-decomposable
are shown in Figure~\ref{fig:Laman}.
In triangle decomposition, the irreducible constituents of the constraint
graph are the minimal constraint graphs,
Definition~\ref{def:minimalCG}.  For the general case, the set of
irreducible constraint graphs has been characterized
in~\cite{Sitharametal} using a network flow approach.  Conceptually,
irreducible constraint graphs must be solved as a single equation
system.  Since the graphs can be arbitrarily complex, so can be the
equation systems.

\begin{figure}
  \begin{center}
   \mbox{\subfigure[]{\includegraphics[scale=0.6]{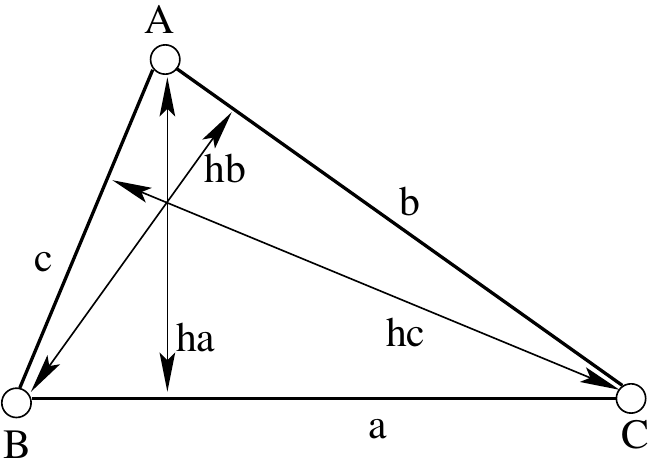}}
         \qquad \qquad
         \subfigure[]{\includegraphics[scale=0.5]{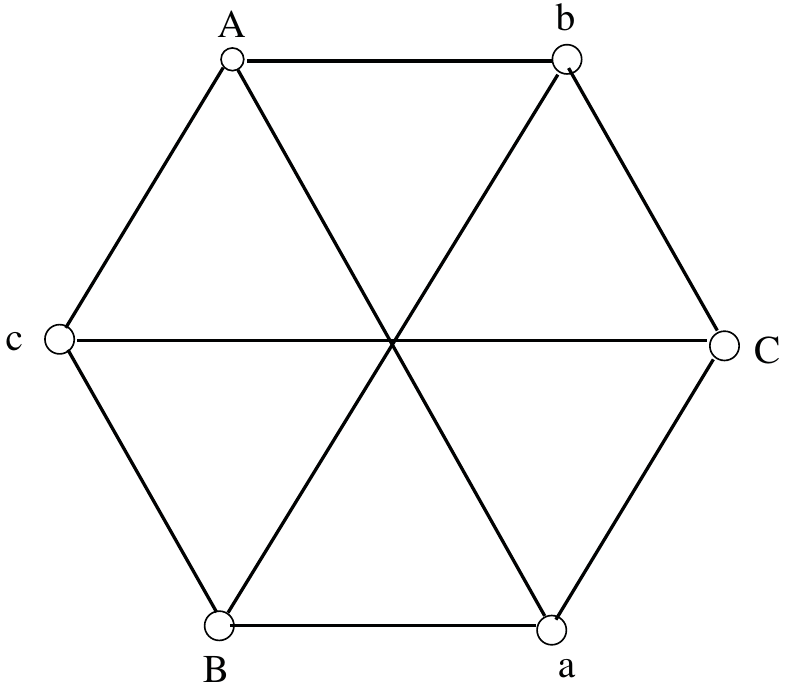}}
   }
  \end{center}
  \caption{Deriving the geometry of a triangle given its three altitudes.
   	   a) Formulating the three altitude problem.
           b) The resulting constraint graph.
  }
  \label{fig:altitudes}
\end{figure}

For the planar case, when we have only points and distances the set of
triangle decomposable graphs coincides with the set of quadratically
solvable graphs. However if we extend to lines and angles, there are
quadratically solvable graphs which are not triangle decomposable.
Consider the problem of finding a triangle from its three altitudes
shown in Figure~\ref{fig:altitudes}(a). The corresponding graph
is shown in Figure~\ref{fig:altitudes}(b) where the hexagon edges are
point-on-line constraints and the diagonals are point-line distance
constraints. The geometric problem is quadratically solvable,
\cite{bib:joan95}, but the graph, $K_{3,3}$, is not triangle
decomposable.

Laman's theorem holds even if we extend the repertoire of geometries to
any geometry having 2 degrees of freedom and the constraints to
virtually any constraint of Table~\ref{tab:constraints}.  However, if
we extend the set of geometries to include for example variable radius
circles, then the Laman condition is no longer sufficient.

\begin{figure}
  \begin{center}
    \includegraphics[scale=0.4]{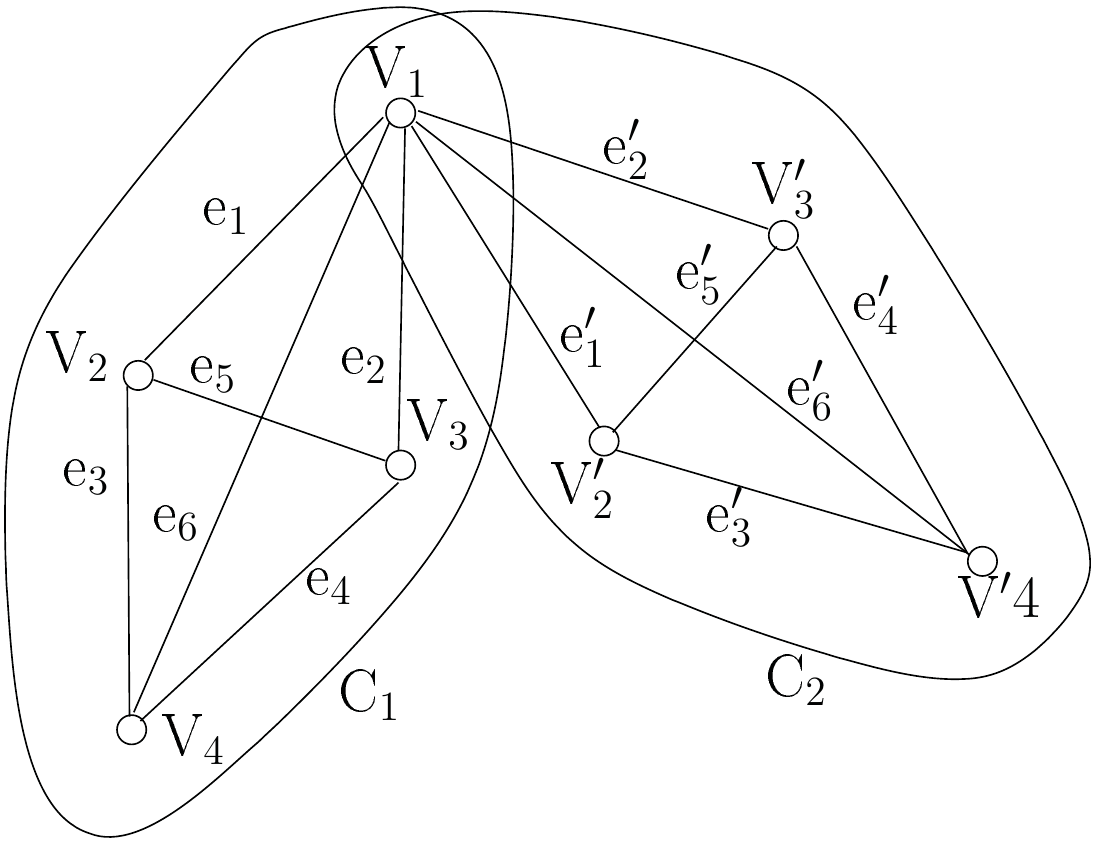}
  \end{center}
  \caption{A Laman graph in 2D that is not rigid.
    V$_1$ is a variable radius circle.
    V$_2$,V$_3$,V$'_2$,V$'_3$ are points.
    V$_4$, V$'_4$ are lines.
    e$_1$,e$_2$,e$'_1$,e$'_2$ are points.
    e$_3$,e$_4$,e$'_3$,e$'_4$ are on constraints.
    e$_5$, e$'_5$ are distances.
    e$_6$, e$'_6$ are distances from cirlce ( circumference).
  }
  \label{fig:not-laman}
\end{figure}

\begin{example}
Consider the GCS
of Figure~\ref{fig:not-laman}.  We have two rigid clusters
$C_1=\{V_1, V_2, V_3, V_4\}$ and $C_2=\{V_1, V'_2, V'_3, V'_4\}$,
where $V_1$ is a variable-radius circle, $V_2, V_3, V'_2, V'_3$ are points,
and $V_4, V'_4$ are lines.
The constraints $e_1,e_2,e'_1,e'_2$ are distances from the center of $V_1$,
$e_6,e'_6$ are distances from the circumference of $V_1$,
$e_5,e'_5$ are distance constraints, and
$e_3,e_4,e'_3,e'_4$ are incidence constraints.
The two clusters share the variable-radius circle $V_1$.

The graph is clearly a Laman graph but is not
rigid, since it is underconstrained --- $C_1$ and $C_2$
can move independently around circle $V_1$.  The problem is also
overconstrained since the radius of $V_1$ can be derived
independently from cluster $C_1$ and from cluster $C_2$.
\end{example}

\subsection{2D and 3D Graphs}
\label{sec:3Dgraphs}

Graph analysis for spatial constraint problems is not nearly as mature
as the planar case.  In the planar case, Definition~\ref{def:minimalCG} conceptualizes
 minimal GCS with which a bottom-up graph analysis would start and a
 top-down analysis would terminate.  Table \ref{tab:minimalGCS2d} lists the configurations.
In 3-space, analogously, the corresponding minimal problem may consist
of three points or three planes, with three constraints
between them.  However, the major configurations shown in Table \ref{tab:minimalGCS3d} also include other cases, for example two lines.

 In the plane, serially adding one additional point or
line requires two constraints.  In 3-space, adding a point or plane
requires three.  The inclusion of lines in 3-space leads to serious
complications.  For example, since lines have 4 dof, see
Table~\ref{tab:dof}, we have to be able to construct lines from given distances from 4
fixed points.  Equivalently, we have to construct a common tangent to
4 fixed spheres, a problem known to have up to 12 real
solutions,~\cite{cmhYuan}.  Work in~\cite{bib:zhou06} has explored the
optimization of the algebraic complexity of 3D subsystems.

\begin{figure}
	\centerline
	{\includegraphics[height=6cm,angle=90]{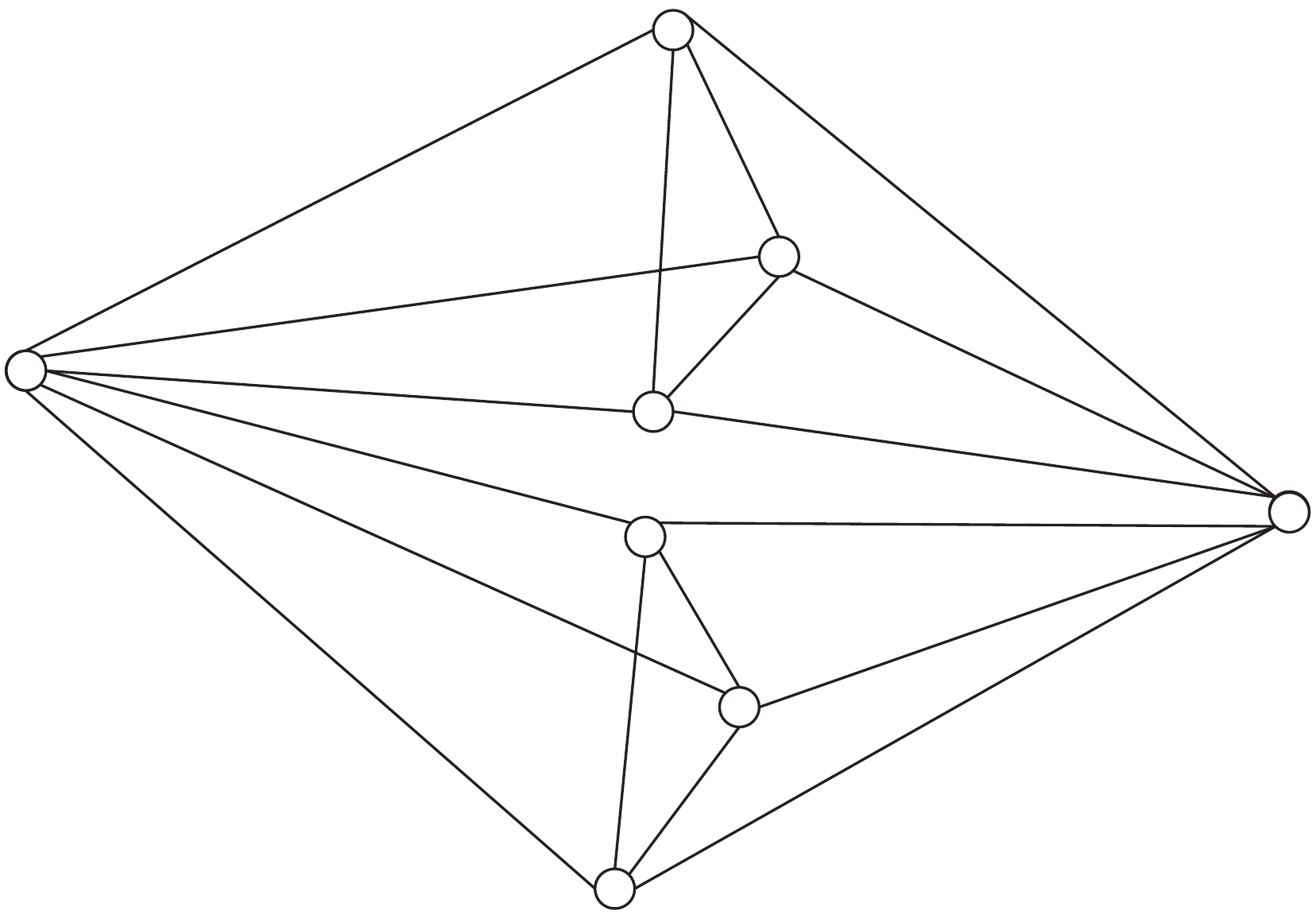}}
	\caption{Two hexahedra sharing two points.}
	\label{fig:bananas}
\end{figure}

In 3D, the Laman condition is not sufficient. Figure \ref{fig:bananas}
illustrates two hexahedra sharing two vertices. If the length of the edges
is given the GCS that arises is also known as the {\em double banana} problem.
The graph is a Laman graph but the problem corresponds to two rigid bodies
(each hexahedron is a rigid body) sharing two vertices and is thus non rigid
in the sense that the two rigid bodies are free to rotate around the axis
defined by the two shared points. The problem is also clearly overconstrained
since the distance of the two shared geometries can be derived
independently by each of the two rigid bodies.

A necessary and sufficient condition for rigidity in 3D has been
recently presented in~\cite{bib:lee2013} for an extended set of
geometries and constraints.  The authors have extended the theory in
\cite{bib:tay84,bib:white87} to characterize systems of rigid bodies
made of points, lines and planes connected by virtually any pairwise
constraint (point-point, point-line, line-plane etc) except
point-point coincidence. In this approach, a multigraph $(V,(B, A))$ is
formulated, where vertices $V$ represent rigid bodies and edges $(B,
A)$ stand for primitive constraints that represent
single equations between the two 6-vectors that describe the rigid body motion
of the two vertices.
Primitive constraints intuitively affect at most
one degree of freedom. Each geometric constraint is translated to a
number of primitive constraints (see Appendix C of \cite{bib:lee2013}).
A distinction is made between primitive
angular and blind constraints: a primitive angular constraint may
affect only a rotational degree of freedom. All other primitive constraints
that may affect either a rotational or translational degree of freedom are called
blind constraints.

Therefore edges are
of two types: angular $(A)$ and blind $(B)$.  Such a scheme is
minimally rigid if and only if there is a subset $B'$ of the blind
edges such that (i) $B-B'$ is an edge disjoint union of 3 spanning
trees and (ii) $A \cup B'$ is an edge disjoint union of 3 spanning
trees.

Note that this characterization does not capture the cases of
Figures \ref{fig:not-laman} and \ref{fig:bananas}. The condition holds for
both cases, but rigidity is not guaranteed since they involve point
coincidences between rigid bodies directly or indirectly.



\section{Solver}\label{sec:CSolver}

After the constraint graph has been analyzed, the implied underlying
equations are to be solved.  We discuss now how to do that.

\subsection{2D Triangle-Decomposable Constraint Problems}

We restrict to points and lines in the Euclidean plane.  As
discussed in Section~\ref{sec:treeDecomp}, triangle-decomposable
constraint systems in the plane require solvers that implement
three operations:
\begin{enumerate}
\item
  The two geometric elements of a minimal subgraph
  (Definition~\ref{def:minimalCG}) are placed consistent with the
  constraint between them.

\item
 A third geometric element is placed by two constraints on two
 geometric elements already placed.

\item
 Given two geometric elements in fixed position, a rigid-body
 transformation is done that repositions the two elements elsewhere.
\end{enumerate}
These operations are applied to the decomposition tree, progressing
from the leaves of the tree to the root.  The solving order is
bottom-up regardless whether the decomposition tree was built
top-down or bottom-up~\cite{bib:owen91,bib:fudos96b,bib:bouma95}.

\begin{example}\label{ex:clusterMerge}
 Consider the constraint system of Example~\ref{ex:truss}.  We choose
 the subgraph induced by $A$ and $B$ as minimal and place $A$ at the
 origin and $B$ on the positive $x$-axis, at the stipulated distance
 from $A$, so executing Operation 1.

 We place $C$ by the two constraints on $A$ and $B$, solving at most
 quadratic, univariate equations, executing Operation 2.  The triangle
 $A, B, C$ is thereby constructed.

 Assume that we have solved the triangle $B, C, D$ in like manner,
 separately and with $B$ and $D$ as vertices of the minimal subgraph.
 We can now assemble the two triangles by a rigid-body transformation
 that moves the triangle $B, C, D$ such that the points $B$ and $C$
 are matched, using Operation 3.
\end{example}
Note that we can extend the geometric vocabulary of points and lines,
adding circles of given radius at no cost.  A fixed-radius circle is
replaced by its center.  A point-on-circle constraint is replaced by a
distance constraint between the point and the center, and a tangency
constraint by a distance constraint between the tangent and the
center.

\subsubsection*{Operation 1: Minimal GCS Placement}

This operation chooses a default coordinate system.  There are three
pairs of geometric elements that occur in minimal GCS: (point, point),
(point, line), and (line, line).  The chosen placements are shown in Table \ref{tab:Op1}.
Other choices could be made.

\begin{table}
\begin{center}
\begin{tabular}{| l | l |}\hline
$d(p_1,p_2)$ & Set $p_1=(0,0)$, $p_2=(d,0)$ \\ \hline
$d(p,L)$     & Place $L$ on the $x$-axis, $p=(0,d)$ \\ \hline
$a(L_1,L_2)$ & Place $L_1$ on the $x$-axis, $L_2$ through the origin at angle $a$ \\ \hline
\end{tabular}
\end{center}
\caption{Placing minimal GCS:  $p$ represents points, $L$ represents lines, $d$ distance, $a$ angle.}
\label{tab:Op1}
\end{table}

\subsubsection*{Operation 2: Constructing one Element from two Constraints}

\begin{figure}
\begin{center}
\mbox{\subfigure[]{\includegraphics[scale=0.5]{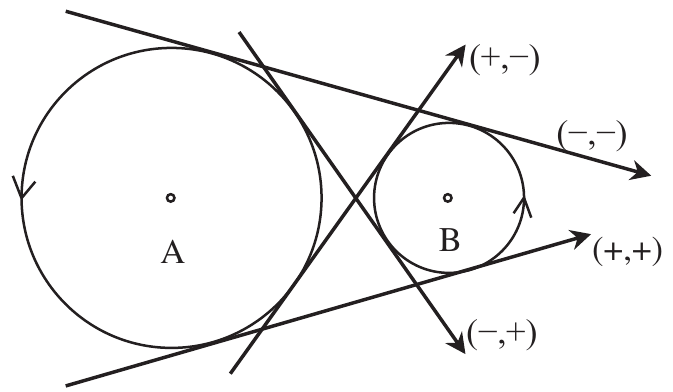}}
      \qquad
      \subfigure[]{\includegraphics[scale=0.5]{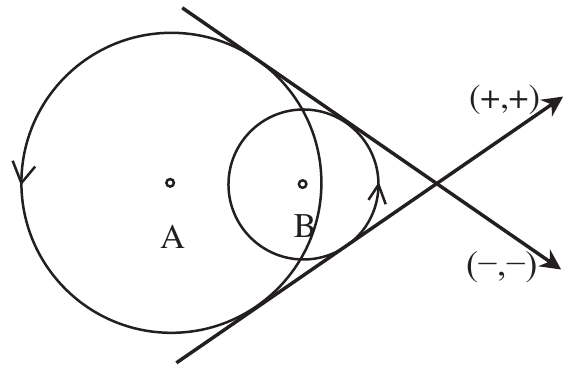}}
      \qquad
      \subfigure[]{\includegraphics[scale=0.5]{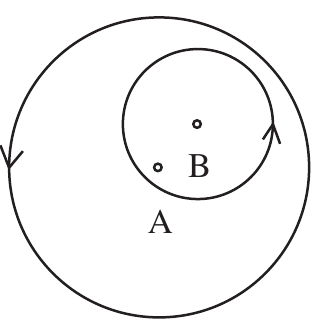}}
     }
\end{center}
\caption{Constructing a line at specific distances from two points.
  Equivalently, finding common tangents of two circles with radius
  equal to the stipulated distances.  The degenerate cases of 1 or 3
  solutions are not shown.}
\label{fig:ppL}
\end{figure}

Two elements $A$ and $B$ are given, a third element $C$ is to be
placed by constraints upon them.  There are six cases, with $p$
denoting a point and $L$ denoting a line.
$$
 \begin{array}{rclrclrcl}
   (p,p)&\rightarrow&p\qquad&(p,L)&\rightarrow&p\qquad&(L,L)&\rightarrow&p\\
   (p,p)&\rightarrow&L\qquad&(p,L)&\rightarrow&L\qquad&(L,L)&\rightarrow&L
 \end{array}
$$
The sixth case, $(L,L)\rightarrow L$ is underconstrained.  See also \cite{bib:bouma95}.
We illustrate with the case
$(p,p)\rightarrow L$.
\begin{example}
In the case $(p,p)\rightarrow L$ a line should be constructed by
respective distance from two points.  Consider the two circles
with the given points as center and radius equal to the stipulated
distance, as shown in Figure~\ref{fig:ppL}.  Depending on the three
distances, there will be 4, 2 or no real solution in general.

Order the points $A$ and $B$, and orient the circles centered at those
points counter-clockwise.  Orient the line to be constructed so that
the projection of $A$ onto the line precedes the projection of $B$.
Then we can distinguish the up to four tangents in a
coordinate-independent way: observe whether the line orientation is
consistent with the circle orientation (+), or whether the
orientations are opposite (--).  See also~\cite{bib:bouma95}.

The degenerate cases where the two circles are tangent to each other
yield three solutions or one. In theses cases there is one double solution that
represents the coincidence of two solutions, with orientations (+,--) and (--,+),
or with orientations (+,+) and (--,--).
\end{example}

\subsubsection*{Operation 3: Matching two Elements}

The operation requires that the geometric elements to be matched be
congruent.  The operation is a rigid-body transformation and is
routine.

\subsection{Root Identification and Order Type}



We noted that constructing one element from
two constraints can have multiple solutions. As a result,
a well constrained GCS has in general an exponential number of
solution instances. We shall illustrate this with the simple
construction $(p,p) \rightarrow p$.

\begin{figure}
  \begin{center}
  \includegraphics[scale=0.4]{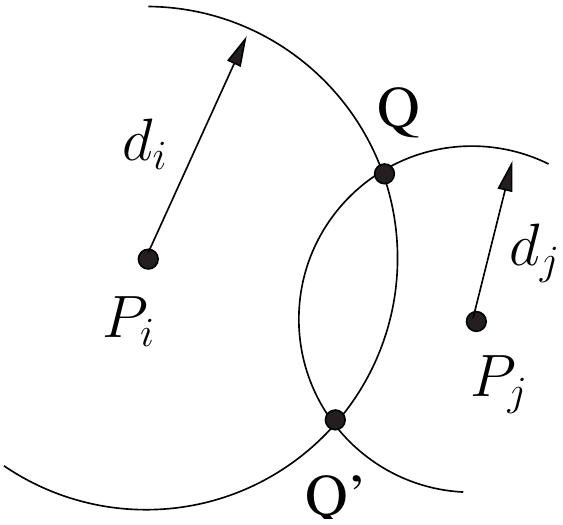}
  \end{center}
  \caption{Placing one point $Q$ from two points $P_i, P_j$
           already known.}
  \label{fig:placep}
\end{figure}

\begin{example}
\label{ex:exponential}
Consider placing $n$ points, by $2n-3$ distance constraints
between them, and assume that the distance constraints are such that
we can place the points by sequentially applying the construction $(p, p)
\rightarrow p$.  In general, each new point can be placed in two
different locations: Let $P_i$ and $P_j$ be known points from which
the new point $Q$ is to be placed, at distance $d_i$ and $d_j$,
respectively. Draw two circles, one about $P_i$ with radius $d_i$, the
other about $P_j$ with radius $d_j$ as shown in
Figure~\ref{fig:placep}. The intersection of the two circles are the
possible locations of $Q$. For $n$ points, therefore, we could have up
to $2^{n-2}$ solution instances.
\end{example}

Note that not all construction paths derive real solutions.  If, in
Example \ref{ex:exponential}, the distance between $P_i$ and $P_j$ is
larger than the sum of $d_i$ and $d_j$, then there is no real solution
for placing $Q$ and therefore any subsequent construction using this
instance of $Q$ is not feasible.  Therefore, one might argue that this
pruning may result in polynomial algorithms.  However, this is
unlikely since the problem of determining whether a well constrained
GCS has a real solution has been shown to be NP-complete
\cite{bib:fudos97}.

In general, an application will require one specific solution,
usually known as the \emph{intended} solution. To identify it
is not always a trivial undertaking. In \cite{bib:bouma95} finding the intended solution is called
the \emph{Root Identification Problem}.
Notice that, on a technical level, selecting the intended solution
corresponds to selecting one among a number of different roots of a
system of nonlinear algebraic equations.

A well constrained GCS would not necessarily include enough
information to identify which solution is the intended one.
Consider the following example.

\begin{figure}
  \begin{center}
  \mbox{\subfigure[]{\includegraphics[scale=0.3]{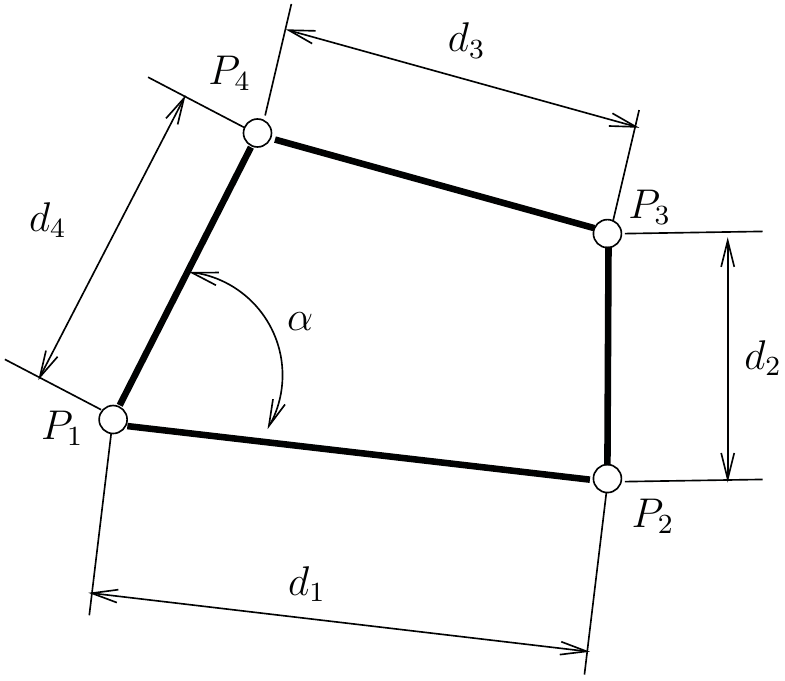}}
        \qquad
        \subfigure[]{\includegraphics[scale=0.3]{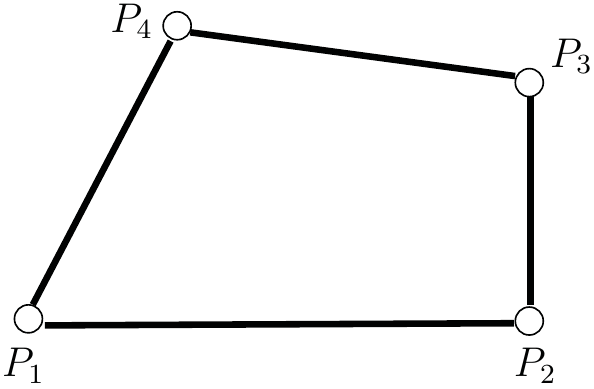}}
        \qquad
        \subfigure[]{\includegraphics[scale=0.3]{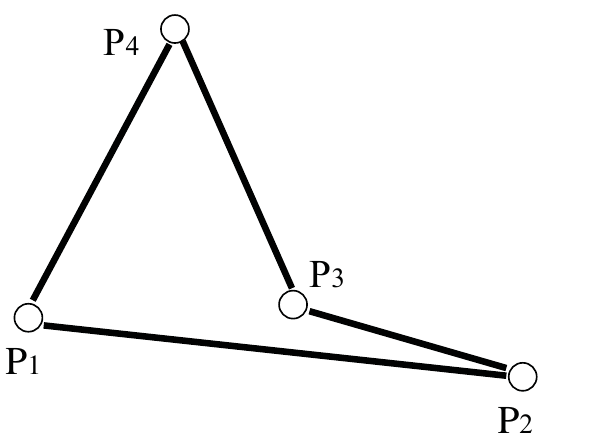}}
        \qquad
        \subfigure[]{\includegraphics[scale=0.3]{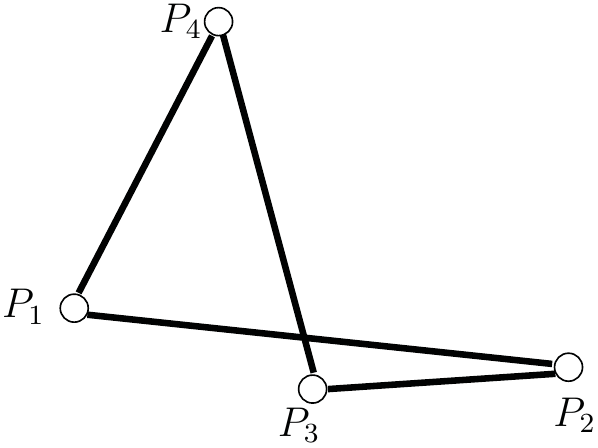}}
  }
  \end{center}
  \caption{a) GCS consisting of four points, four straight segments,
              four point-point distances and an angle.
           b),c) and d) Three different solution instances to the GCS.
          }
  \label{fig:quadri}
\end{figure}

\begin{example}
\label{exa:rip_a}
The well constrained GCS in Figure~\ref{fig:quadri}a consists of
four points, four straight segments, four point-point distances and an
angle.
The solution includes four instances. Two
correspond to the one shown in Figure~\ref{fig:quadri}b and to a
symmetric arrangement of the same shape. Solution instances in
Figures~\ref{fig:quadri}c and~\ref{fig:quadri}d are structurally
different.
\end{example}

Clearly, the GCS sketch in Figure~\ref{fig:quadri}a does not include
any hint on which solution instance must be chosen to be displayed on
the user's screen. Thus, additional information must be supplied to
the solver. In~\cite{bib:bouma95}, approaches applied to
overcome this issue have been classified into five categories:
Selectively moving geometric elements, adding extra constraints to
narrow down the number of possible solution instances, placement
heuristics, a dialogue with the constraint solver that identifies
interactively the intended solution, and applying a design paradigm
approach based on structuring the GCS hierarchically. Next we
elaborate on each category.

\subsubsection*{Moving selected geometry}

In this approach, a solution
is presented graphically to the user.  The user
selects, again graphically, certain geometric elements that are considered misplaced.
The user  then moves those elements where they should
be placed in relation to other elements.
This approach is used by the DCM
solver,~\cite{bib:dcubed,bib:owen91}.

\subsubsection*{Adding extra constraints}

Adding a set of extra constraints to narrow down which is the intended
solution instance is an intuitive and simple approach to solving the
root identification problem. Extra constraints could capture domain
knowledge from the application or could be just geometric --- and
actually over-specify the GCS.  Unfortunately, both ideas result in NP-complete
problems~\cite{bib:bouma95}.  Nevertheless, extra constraints along with
genetic algorithms have been applied to solve the root identification
problem showing a promising potential.
Authors in \cite{bib:yeguas11} argue that the approach is both
effective and efficient in search spaces with up to $2^{100}$ solution
instances.
A different application of genetic algorithms to solve the root
identification problem is described in~\cite{bib:cao04}. The approach
mixes a genetic algorithm with a chaos optimization method.

\begin{example}
Genetic algorithms described
in~\cite{bib:joanarinyo02g2,bib:luzon05,bib:joanarinyo09}, use
extra geometric and topological constraints defined as
logical predicates on oriented geometries. For example,
assume that the polygon in Figure~\ref{fig:quadri}a is
oriented counterclockwise. The solution shown in
Figure~\ref{fig:quadri}c  would be selected as the intended one
by requiring the following two predicates to be fulfilled
$$PointOnSide(P_3, \stackrel{\longrightarrow}{P_1P_2}, left), \qquad
Chirality(\stackrel{\longrightarrow}{P_2P_3},
          \stackrel{\longrightarrow}{P_3P_4}, cw)$$
\end{example}

\subsubsection*{Order-based heuristics}

\begin{figure}
  \centering
  \mbox{\subfigure[]{\includegraphics[scale=0.3]
                     {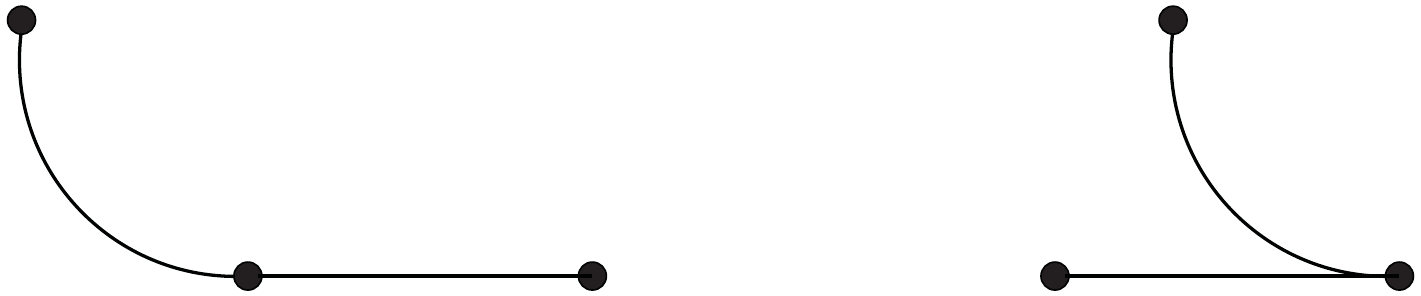}}
        \qquad
        \subfigure[]{\includegraphics[scale=0.2]
                     {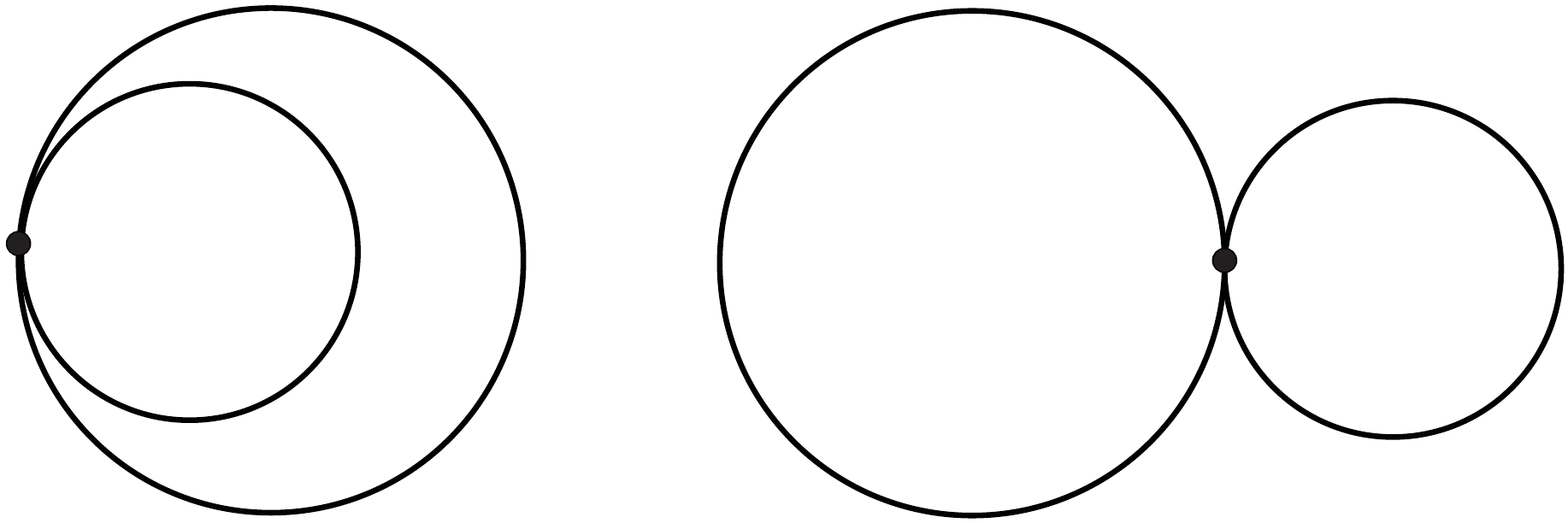}}
  }
  \caption{Tangency types.
           a) Arc and segment tangency.
           b) Circle-circle tangency.
          }
  \label{fig:tangency}
\end{figure}

All solvers known to us derive information from the initial GCS sketch
and use it to select a specific solution. This is reasonable, because
one can expect that a user sketch is similar to what is intended.  For
instance, by observing on which side of an oriented line a specific
point lies in the input sketch it is often appropriate to select
solutions that preserve this sidedness.  The solver described
in~\cite{bib:bouma95} seeks to preserve the sidedness of the geometric
elements in each construction step: the orientation of three points
with respect to each other, of two lines and one point, and of one
line and two points.
The work described in \cite{bib:bouma95} implements an additional
heuristic for arc 
tangency which aims at preserving the type of tangency present in the
sketch.  See Figure~\ref{fig:tangency}. When the rules fail, the
solver opens a dialogue to allow the user to amend the rules as the situation
might require. These heuristics are also applied in the solver
described in~\cite{bib:joanarinyo97a}.

\begin{figure}
  \centering
  \mbox{\subfigure[]{\raisebox{4mm}{\includegraphics[scale=0.3]
                                    {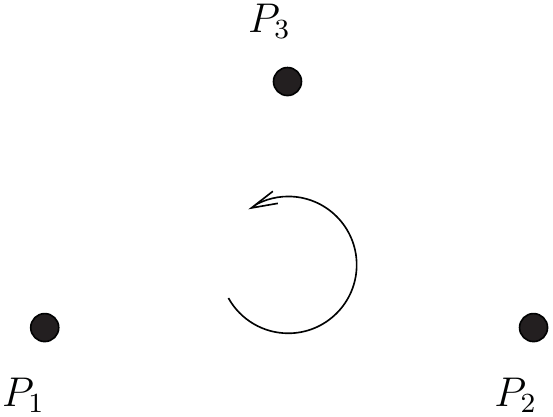}}}
        \hspace{2cm}
        \subfigure[]{\includegraphics[scale=0.3]
                    {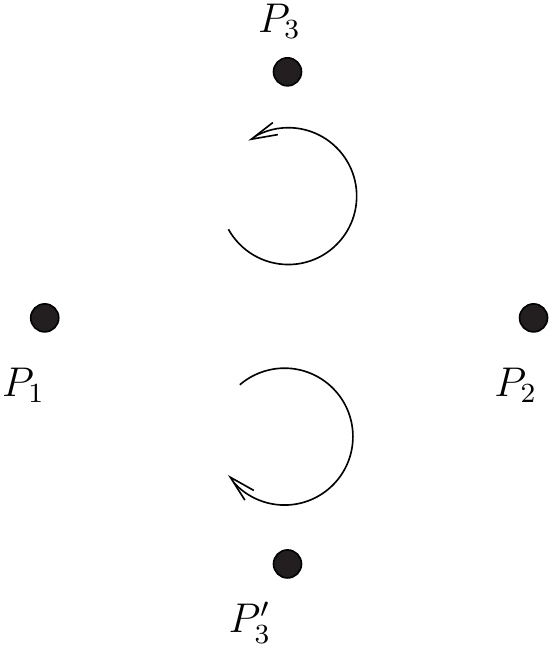}}
  }
  \caption{Placing three points, $P_1$, $P_2$ and $P_3$,
           relative to each other.
           a) Points placed in the initial sketch and induced orientation.
           b) $P_3$ and $P'_3$ are two possible placements for the
           third point. Preserving the orientation defined in the
           sketch leads to select $P_3$ as the intended placement.}
  \label{fig:threepoints}
\end{figure}

\begin{example}
Consider placing three points, $P_1$, $P_2$ and $P_3$, relative to
each other. The points have been drawn in the initial sketch in the
position shown in Figure~\ref{fig:threepoints}a. The order
defined by the points can be determined as follows. Determine where
$P_3$ lies with respect to the line $\stackrel{\longrightarrow}{P_1P_2}$.
If $P_3$ is on the line, then determine whether
it lies between $P_1$ and $P_2$, preceding $P_1$ or following
$P_2$. The solver will preserve this orientation if possible. For this
example, the solver will choose the point $P_3$ as shown
in Figure~\ref{fig:threepoints}b.
\end{example}

\subsubsection*{Dialog with the solver}

A useful paradigm for user-solver interaction has to be intuitive and
must account for the fact that most application users will not be
intimately knowledgeable about the technical working of the solver. So,
we need a simple but effective communication paradigm by which the
user can interact with the solver and direct it to a different
solution, or even browse through a subset of solutions in case the one
shown in the user's screen is not the intended one.

\begin{figure}
  \centering
  \vbox{
  \mbox{\subfigure[]{\includegraphics[scale=0.2]{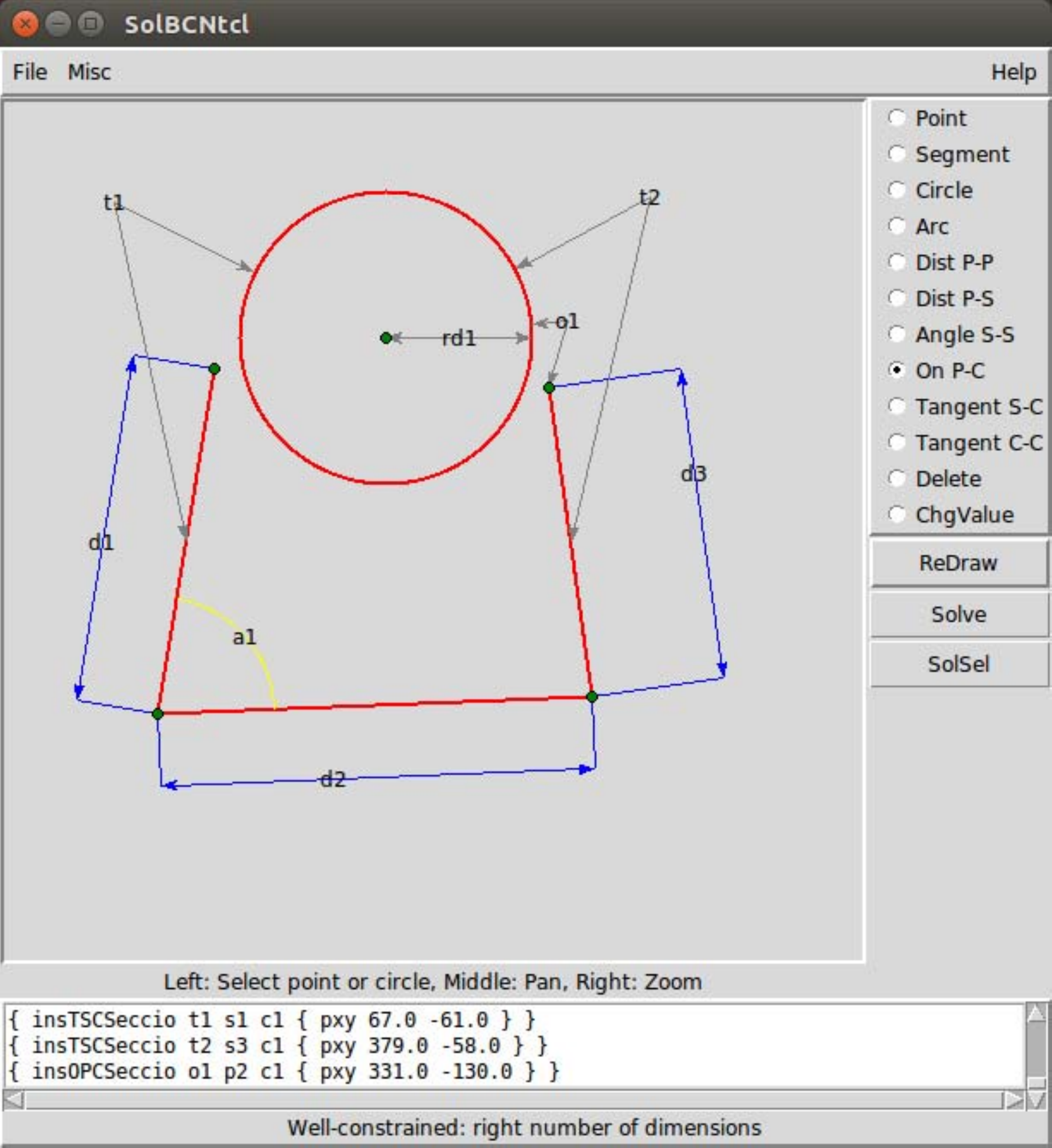}}
        \hspace{0.9cm}
        \subfigure[]{\includegraphics[scale=0.2]{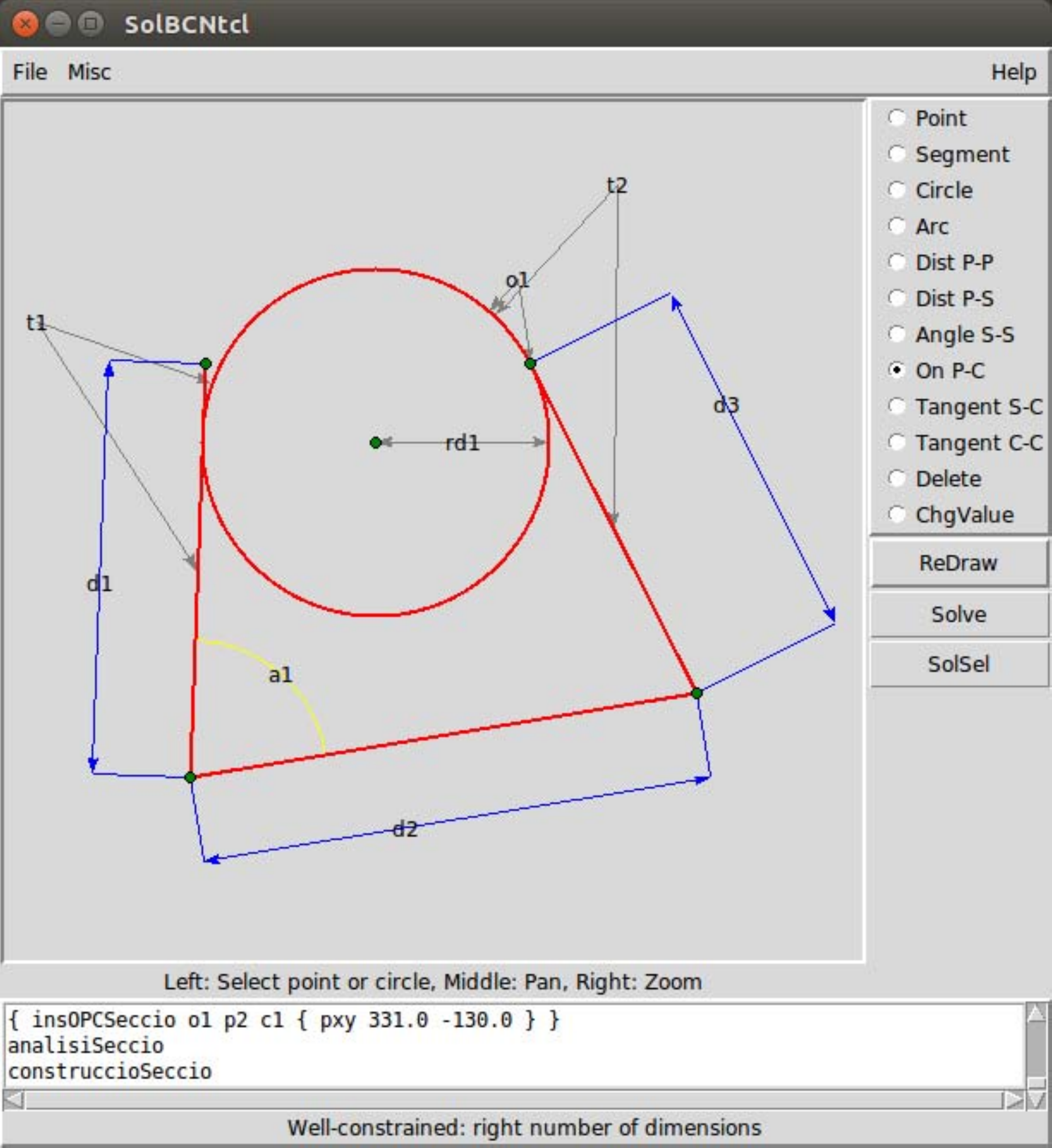}}
  }

  \mbox{\subfigure[]{\raisebox{3mm}{\includegraphics[scale=0.6]
                                    {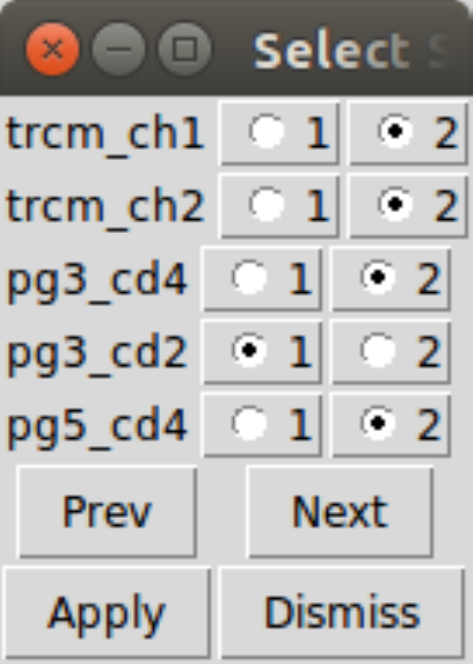}}}
        \hspace{2cm}
        \subfigure[]{\includegraphics[scale=0.2]{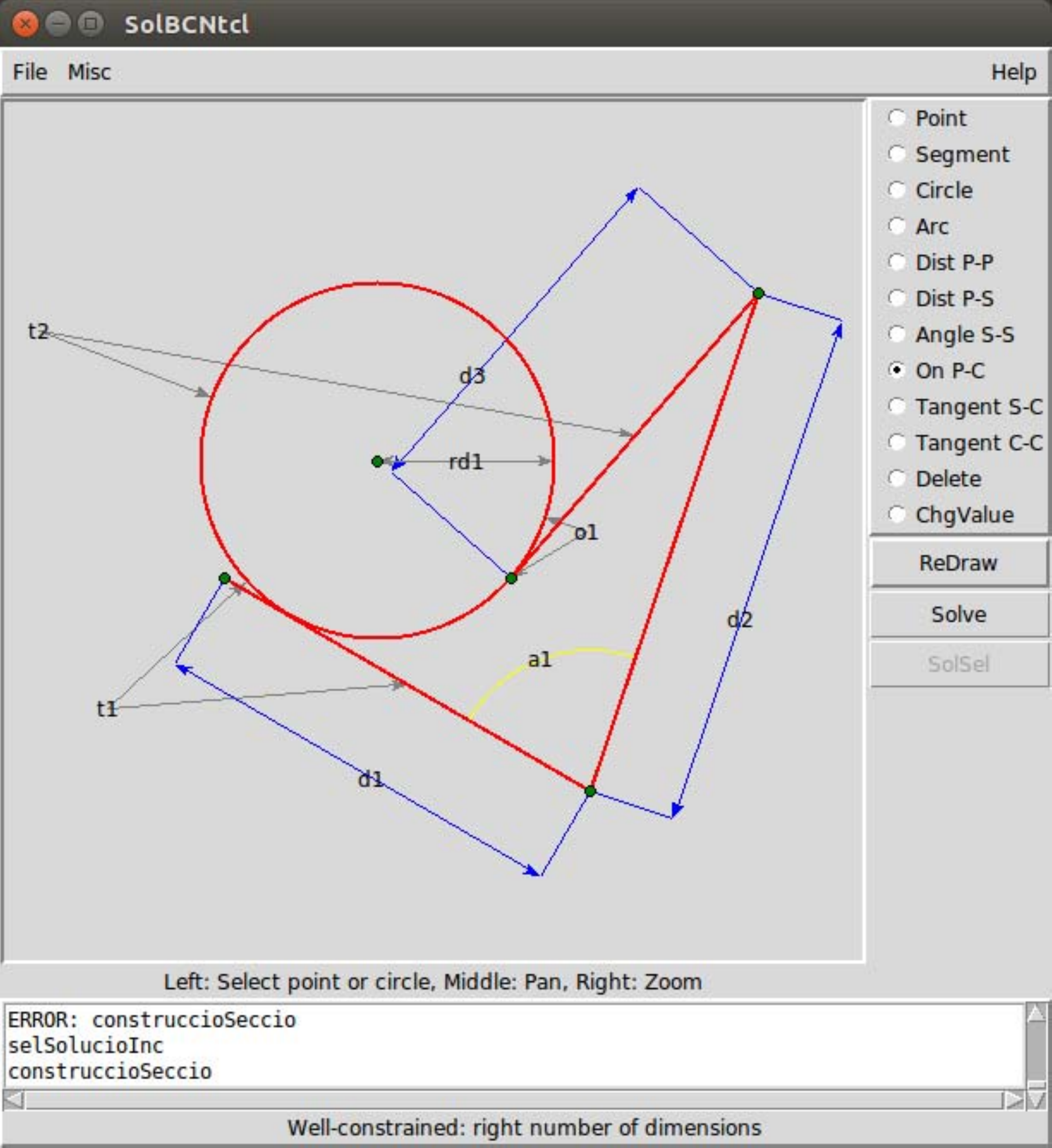}}
  }
  }
  \caption{User-solver dialog offered by the ruler-and-compass solver
           described in~\cite{bib:joanarinyo97a}.
           a)~GCS sketch.
           b)~Solution instance selected by the heuristics
              implemented in the solver.
           c)~Solution instance selector.
           d)~Solution instance displayed after changing the
              square root signs of some construction stpdf.
           }
  \label{fig:selectorSolBCN}
\end{figure}

\begin{example}
\label{exa:solBCN}
SolBCN, a ruler-and-compass solver described in~\cite{bib:joanarinyo97a}
offers a simple user-solver interaction tool.
Figures~\ref{fig:selectorSolBCN}a and b respectively show the GCS sketch
input by the user and the solution instance selected by the heuristic
rules implemented. Then, the user can trigger the
solution selector by clicking on a button and a list with the set of
construction stpdf where a quadratic equation must be solved is
displayed,  Figure~\ref{fig:selectorSolBCN}c. The user can then change
the square root sign for each of these construction stpdf by  either
selecting it directly or navigating with the next/previous pair of buttons.
Figure~\ref{fig:selectorSolBCN}d shows a solution different from the
first one so obtained.
\end{example}

Navigating the GCS solution space using the approach illustrated in the
Example~\ref{exa:solBCN} is simple.  But it has obvious drawbacks. On the
one hand, the number of items in the list selector grows exponentially
with the number of quadratic construction stpdf in the GCS. On
the other hand it is difficult to anticipate how choosing a root sign
for a construction step will affect the solution selected by the next
sign chosen by the user.

These problems are avoided by considering that, conceptually, all
possible solution instances of a GCS can be
arranged in a tree whose leaves are the different instances, and whose
internal nodes correspond to stages in the placement of individual
geometric elements. The different branches from a particular node are
the different choices for placing the geometric element. Since the
tree has depth proportional to the number of elements in the GCS,
stepping from one solution instance to another is proportional
to the tree depth only. Moreover, it is possible to define an
incremental approach by allowing the user to select at each
construction step which tree branch should be used. In solvers based on
the DR-planning paradigm, \cite{bib:hoffmann01a}, this tree naturally is
the construction plan generated by the solver.

\begin{figure}
  \centering
  \vbox{
  \mbox{\subfigure[]{\includegraphics[scale=0.5]{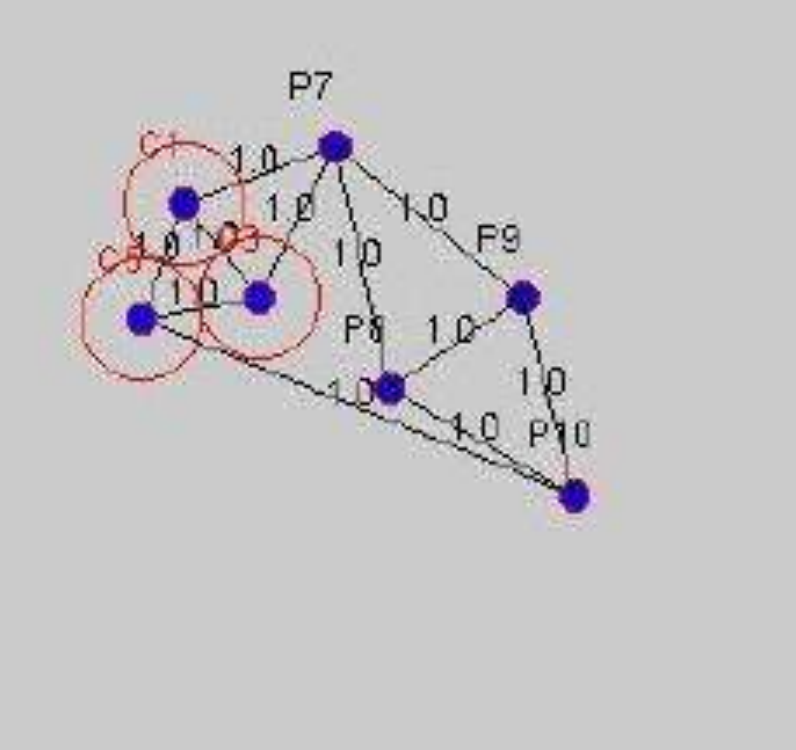}}
        \subfigure[]{\includegraphics[scale=0.5]{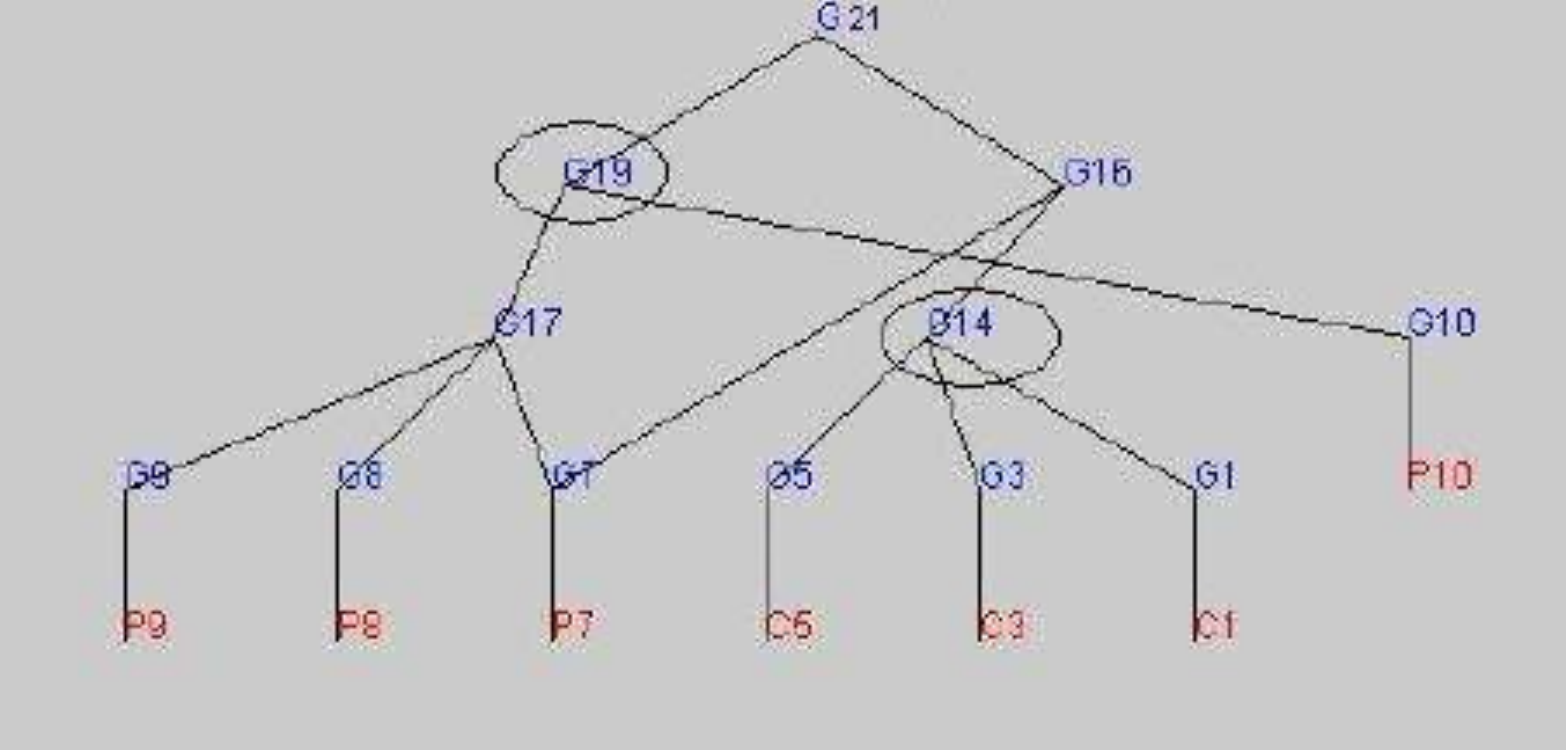}}
        }
  \mbox{\subfigure[]{\includegraphics[scale=0.5]{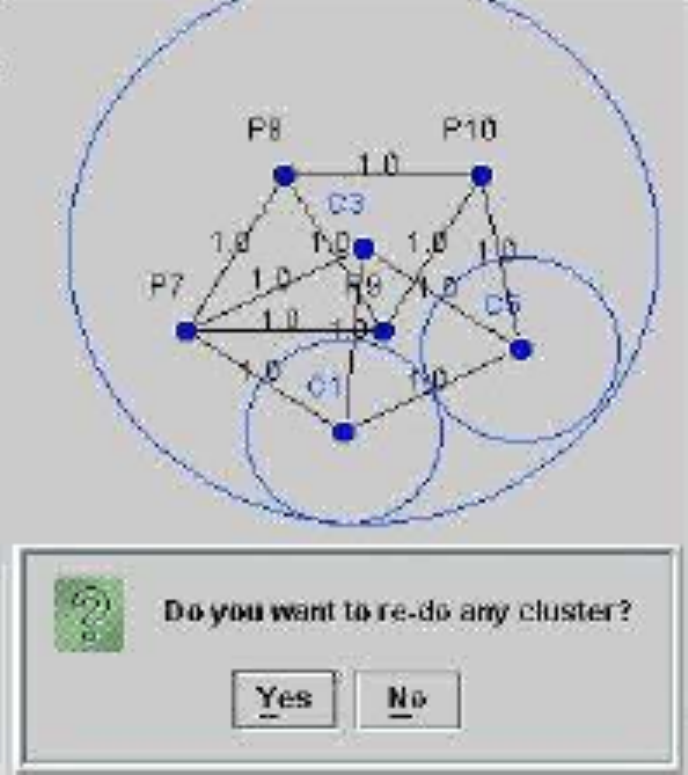}}
        \qquad
        \subfigure[]{\includegraphics[scale=0.5]{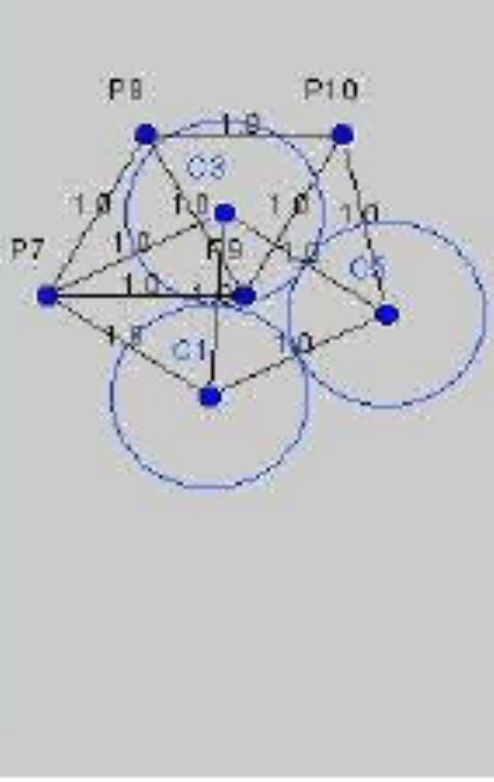}}
       }
  }
  \caption{Incremental solution space navigation described
           in~\cite{bib:sitharam06}.
           a) GCS problem including three circles.
           b) Construction plan graph for the GCS solution.
           c) GCS solution instance after choosing
              one of the possible solutions for each construction step.
           d) A different GCS solution instance after
              rebuilding the partial construction corresponding to the
              construction step labeled G14 in the tree.
          }
  \label{fig:ESMselector}
\end{figure}

\begin{example}
The Equation and Solution Manager~\cite{bib:sitharam06},
features a scalable method to navigate the solution space of GCS.
The method incrementally assembles the desired solution to the GCS and
avoids combinatorial explosion, by offering the user a visual
walk-through of the solution instances to recursively constructed
subsystems and by permitting the user to make gradual, adaptive
solution choices. Figure~\ref{fig:ESMselector} illustrates the approach.
\end{example}

\subsubsection*{Design paradigm approach}

One of the difficulties in selecting the intended solution of a GCS
stems from the fact that geometric elements in a problem
sketch are not grouped into logical structures.
Authors in \cite{bib:bouma95} argue that hierarchically structuring
the constraint problem would alleviate the complexity of solving the
root identification problem, for example grouping geometric elements
as design features.  First a basic, dimension-driven sketch would be
given. Then, subsequent dimension-driven stpdf would modify the
basic sketch and add complexity. By doing so, the
design intent would become more evident and some of the
technical problems would be simplified.

\begin{figure}
  \centering
  \mbox{\subfigure[]{\includegraphics[scale=0.525]
                                 {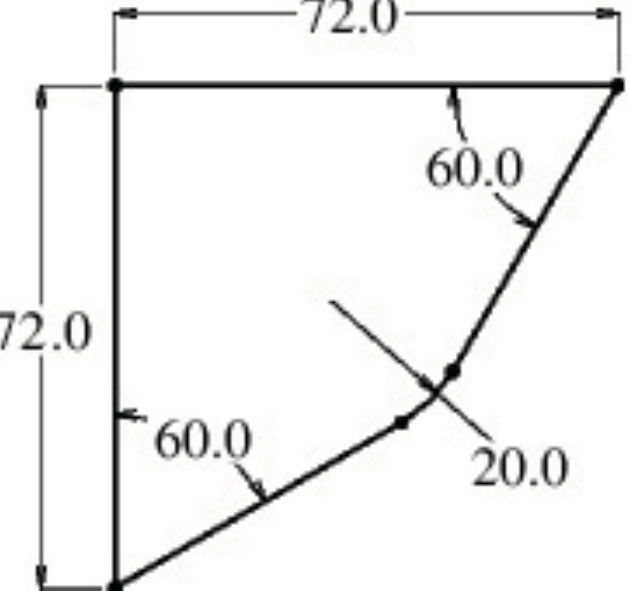}}
        \qquad
        \subfigure[]{\includegraphics[scale=0.5]
                                  {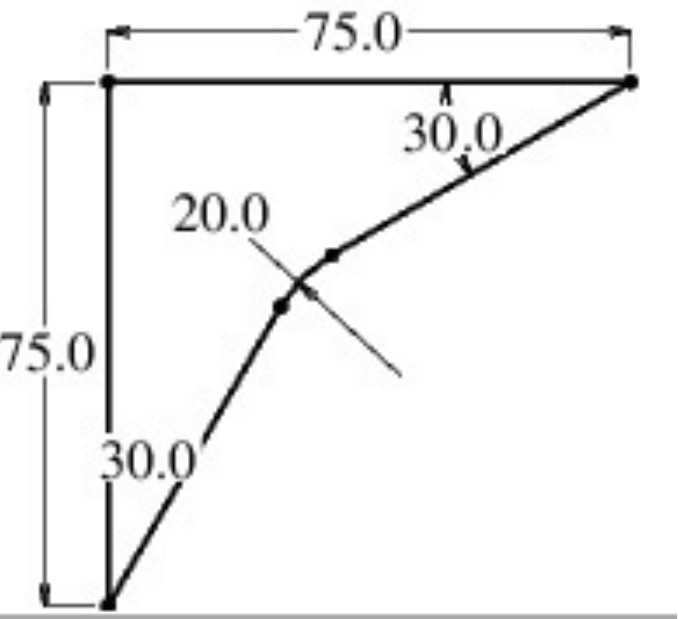}}
        \quad
        \subfigure[]{\includegraphics[scale=0.5]
                                  {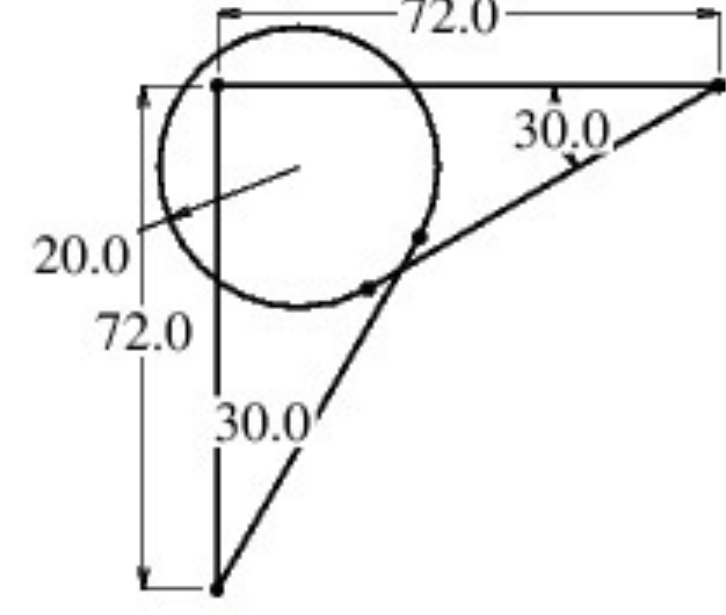}}
  }
  \caption{Solution selection by the design paradigm approach:
           Panel (a) shows the final GCS, panels
           (b) and (c) two different solution instances.
           If the arc is introduced as a rounding feature of a constrained quadrilateral,
           then selecting solution (b) over solution (c) is a logical choice.}
  \label{fig:paradigm}
\end{figure}

\begin{example}
Consider solving the GCS in Figure~\ref{fig:paradigm}a.
The role of the arc is clearly to round the adjacent segments, and
thus it is most likely that the solution shown in
Figure~\ref{fig:paradigm}b is the one the user meant rather than the
one in Figure~\ref{fig:paradigm}c, when changing the angles to
$30^{\circ}$.  However, the solver would be unaware of the intended meaning of
the arc.
Instead, the user could sketch first the quadrilateral without the arc, and
then add the arc to round a vertex.
When changing some of the dimensional constraints,
the role of the arc would remain that of a round, so preserving the user intent.
\end{example}

\subsection{Extended Geometric Vocabulary}


So far we discussed 2D constraint solvers that use only points and
lines, as well as circles of given radii.  Now we will add other geometric element types.  This has
implications for both the equation solvers as well as for the
constraint graph analysis.

The simplest addition allows GCS to include geometric elements of the
new type, but only if these elements can be constructed sequentially,
from explicit constraints on a set of already placed geometric
elements.  A more difficult addition would allow the use of elements
of the added type in the same way as the core vocabulary of points and
lines.  Note that a new element type can have more than two degrees of
freedom.

\subsubsection*{Variable-Radius Circles}\label{sec:varraded}

Circles whose radii have not been given explicitly are arguably the
most basic extension of the 2D core solver.  Variable-radius circles
have three degrees of freedom.  We consider two ways in which they can arise:
\begin{enumerate}
\item
 A variable-radius circle is to be constructed by a sequential step from three, already placed geometric entities.
\item
 A variable-radius circle is to be determined in a step analogous to merging three clusters (Definition \ref{def:treeDecomp}) and the circle acts as a cluster.
\end{enumerate}
Note that a variable-radius circle cannot be a shared element in the
sense of Definition~\ref{def:treeDecomp}. Shared elements have already
been constructed; therefore a shared variable-radius circle has
already become a fixed-radius circle.

In the following, we consider points to be circles of zero radius.
Consequently, there are four ways in which a variable-radius circle
can be constructed sequentially.  Table~\ref{tab:VCsequential}
summarizes the equation systems.

It is advantageous to convert a line-distance constraint into two
separate tangency constraints, so simplifying the equations that must
be solved.  For example, if the sought circle $C$ is to be at distance
$d$ from line $L$, we work instead with two problems, one in which the
circle $C$ is to be tangent to a parallel line $L_+$ of $L$. Here
$L_+$ is at distance $d$ from $L$, on one side of $L$.  In the other
problem, $C$ is to be tangent to a parallel line $L_-$ also at
distance $d$, but on the other side of $L$.  Analogously, perimeter
distance from a given circle reduces to tangency with a circle whose
radius has been increased or reduced by said distance.

When deriving the algebraic equations, we work with
\emph{cyclographic maps}, orienting both lines and circles.  Briefly,
an oriented circle $C$ is mapped to a {\em normal cone} $\mu(C)$, with
an axis parallel to the $z$-axis and a half angle of $\pi / 4$.  The
cone intersects the $xy$-plane in $C$.  Depending on the orientation,
the apex of the cone is above or below the $xy$-plane.
Considered as zero-radius circle, the point $P$ in the $xy$-plane maps
to the normal cone $\mu(P)$ with apex in the $xy$-plane.  Oriented
lines $L$ in the $xy$-plane are mapped to planes through $L$ at an
angle of $\pi / 4$ with the $xy$-plane.  This reformulates the
constraint problem as a spatial intersection problem of planes and
cones. See
\cite{bib:Pottmann98,bib:cmhEG92,bib:hoffmann02a,bib:hoffmann02b,bib:chiang04}
for details and further reading.

The intersection of two normal cones is a conic, in affine space, plus a shared circle at infinity.
The conic lies in a plane whose equation is readily obtained by subtracting the two cone equations: if $K_1=0$ and $K_2=0$ are the two normal cones, then $K_1 - K_2 = 0$ is the plane that contains the conic in which the two cones intersect.

\begin{example}\label{ex:Apollo}
Consider the sequential construction problem of finding a circle that
is tangent to three given circles.  This is the classical Apollonius
problem that has eight solutions in general.

We orient the circles and require that the sought circle be oriented
consistently with the given circles at the points of tangency.  After
orienting the given circles, we can map the problem to the
intersection, in 3-space, of three normal cones, $C_1$, $C_2$, and
$C_3$, each arising from an oriented circle.  Intersect two cone
pairs, say $C_1\cap C_2$ and $C_1\cap C_3$, obtaining two planes that,
in turn, intersect in a line $L$ in 3-space.  Then intersect $L$ with one of the
cones, say $C_2$.  Two points are obtained that, understood as the
apex of a normal cone, map each to one (oriented) circle in the plane
that is a solution; see \cite{bib:ramanathan96}.  Algebraically, the
solution is obtained by solving linear equations plus one univariate
quadratic equation.

There are 8 ways to orient the three circles, but they correspond
pairwise, so only four such problems must be solved.  If one or more
circles are points, they must be considered oriented both ways.  So,
for each zero-radius circle, the number of solutions reduces by a
factor of 2.  The special cases of the Apollonius problem have been
mapped out and solved in~\cite{bib:ramanathan96} using cyclographic
maps.
\end{example}

\begin{table}
\begin{center}
\begin{tabular}{|c|c|l|}
\hline
\rule{0pt}{12pt}Given & Equation & \rule{1cm}{0pt}Notes \\
Elements & System & ~ \\
\hline
\hline
\rule{0pt}{12pt}$LLL$ & (1,1) & intersect two angle bisectors \\
$LLC$ & (1,2) & intersect two planes and a cone \\
$LCC$ & (1,1,2) & intersect two planes and a cone \\
$CCC$ & (1,1,2) & Apollonius problem; Example \ref{ex:Apollo} \\
\hline
\end{tabular}
\end{center}
\caption{Sequential construction of variable-radius circles.  All
  constraints are tangent constraints.  Equations formulated using
  cyclographic maps.  For the cases $LCC$ and $CCC$ the linear
  equation(s) are from intersecting two
  cones.}\label{tab:VCsequential}
\end{table}

Now consider the determination of variable-radius circles in a cluster
merge.  Here, there are two constraints from each cluster to the
variable-radius circle to be constructed, and the two clusters share a
geometric element.  The situation is analogous to the triangle merge
characterized in Definition~\ref{def:treeDecomp}.  The various cases
and how to solve them have been studied and solved in
\cite{bib:hoffmann02a,bib:hoffmann02b,bib:chiang04}.  Specifically,
\cite{bib:hoffmann02a,bib:hoffmann02b} map out the cases in which the
constraints are on the perimeter of the variable-radius circle; and
\cite{bib:chiang04} considers constraints on the center of the
variable-radius circle as well.

Table~\ref{tab:VCCtransrot} and
\ref{tab:VCCcenter} summarize the results from those papers.  The
approach is conceptually as follows.  Let $S_1 = \{E_0 , E_1 , E_2 \}$
and $S_2 = \{E_0 , E_3 , E_4\}$ be the clusters constraining the
variable-radius circle.  The two clusters share element $E_0$, either
a line denoted $L$, or a circle denoted $C$.  The clusters can move
relative to each other, translating along the shared line if $E_0 =
L$, or rotating about the center of the shared circle if $E_0 = C$.
Elements $E_1$ and $E_2$ belong to $S_1$ and constrain the sought
circle.  Likewise, $E_3$ and $E_4$ are the constraining elements of
$S_2$.  Proceed as follows:
\begin{enumerate}
\item
  Fix the cluster $S_1$ that has the more difficult constraining
  elements $E_1$ and $E_2$; i.e., the cluster with the larger number
  of circles.

\item
  Choose a convenient coordinate system: the shared line $E_0=L$ as
  the $x$-axis, or the origin as the center of the shared circle if
  $E_0=C$.

\item
  Construct the cyclographic map of all constraining elements.  The
  cones and planes of $S_2$ are parameterized by the distance $d$
  between $S_1$ and $S_2$, or else by the angle $\theta$ between $S_1$
  and $S_2$.

\item
  Construct three planes from the constraining elements $E_1,\ldots, E_4$.
  They are either cyclographic maps of lines, or normal cone
  intersections.  The elements of $S_2$ give rise to parameterized
  coefficients, by distance $d$ for translation along the $x$-axis, or
  by angle $\theta$ for rotation around the origin, of the moving
  cluster $S_2$.  Intersect the planes, so obtaining a point with
  parameterized coordinates.

\item
  Substitute the parameterized point into the equation of the element
  $E_1$ of the fixed cluster, so obtaining a univariate polynomial
  that finds the intersection point(s) of the four cyclographic
  objects; a polynomial in $d$ or $\theta$.

\item
  Solve the polynomial as described, each obtained by a
  particular configuration of orientations.
\end{enumerate}


\begin{table}
\begin{center}
\begin{tabular}{|c| l |c|}
\hline
Constraint type & Planes & Polynomial degree
\rule{0pt}{12pt} \\
\hline\hline
$E(LL,LL)$ & $[L_2],~ [L_3]^t,~ [L_4]^t$ & (1,2) \rule{0pt}{12pt}\\
\hline
$E(CL,LL)$ & $[L_2],~ [L_3]^t,~ [L_4]^t$ & (2,4) \rule{0pt}{12pt}\\
\hline
$E(CL,CL)$ & $[L_2],~ [L_4]^t,~ ([C_1], [C_3]^t)$ & (4,4) \rule{0pt}{12pt}\\
\hline
$E(CC,LL)$ & $([C_1],[C_2]),~ [L_3]^t,~ [L_4]^t$ & (2,4) \rule{0pt}{12pt}\\
\hline
           & $([C_1],[C_2])$   & \rule{0pt}{12pt}\\
$E(CC,CL)$ & $([C_1],[C_3]^t)$ & (4,4) \\
           & $[L_4]$ &\\
\hline
           & $([C_1],[C_2])$   &\rule{0pt}{12pt}\\
$E(CC,CC)$ & $([C_1],[C_3]^t)$ & (4,4) \\
           & $([C_3],[C_4])^t$ & \\
\hline
\end{tabular}
\end{center}
\caption{Cluster cases; all constraints on circle perimeter.  $E(...)$ denotes whether clusters share a line $L$, the translational case, or share a circle $C$, rotational case.  $[X]$ denotes the cyclographic map equation  $\mu(X)$ of $X$; $[X]^t$ denotes the equation with  coefficients parameterized by distance $t$ (translation case) or by angle $\theta$ (rotation case).  $(X,Y)$ denotes the intersection plane equation of $X$ and $Y$.  The parameterized point is substituted into the equation $[C_1]$, except for the first case where it is substituted into $[L_1]$.  $(m,n)$ denotes the equation degrees, namely $m$ for the translation case $E=L$,
 and $n$ for the rotation case $E=C$.}
\label{tab:VCCtransrot}
\end{table}

Some of the constraints can be on the center $c$ of the
variable-radius circle, and~\cite{bib:chiang04} considers those cases.
Note that there can be at most two constraints on the center of the
variable-radius circle, for otherwise the relative position of $S_2$
to $S_1$ would be determined and the role of the variable-radius
circle would be curtailed.

The problem is again solved in the same conceptual manner, but with a
twist.  When a constraint is placed on the center $c$ of the
variable-radius circle, the constraint can be expressed by extending
cyclographic maps with a map $\tau(X)$.  Here, $\tau(L)$ is a vertical
plane through the line $L$.  Moreover,  $\tau(C)$ is a cylinder through the circle $C$
with axis parallel to the $z$-axis.  The results so obtain are
summarized in Table~\ref{tab:VCCcenter}.

A problem is denoted $E_0 (E_1 E_2 , E_3 E_4)$.  $E_0$ is the shared
element by the two clusters, a line $L$ or a circle $C$.  $E_1$ and
$E_2$ are the two elements of the fixed cluster constraining the
variable-radius circle.  $E_3$ and $E_4$ are the two elements of the
moving cluster constraining the variable-radius circle.  The numbers
(m,n) are the equation degrees when $E_0=L$ (m), and $E_0=C$ (n).  An
element $E'_k$ constrains the center, an element $E$ the
circumference, of the variable-radius circle.

\begin{table}
\begin{center}
\begin{tabular}{|lr|lr|}
\hline
\multicolumn{2}{|c|}{One center constraint} &
\multicolumn{2}{c|}{Two center constraints\rule{0pt}{12pt}} \\
Problem & Degree & Problem & Degree\rule{0pt}{12pt}\\
\hline\hline
E(LL,LL') & (1,2) & E(LL,L'L') & (1,2)\rule{0pt}{12pt}\\
         &       & E(LL',LL') & (1,2) \\
\hline
E(CL,LL') & (2,4)    & E(CL,L'L') & (2,4)\rule{0pt}{12pt}\\
E(CL',LL) & (2,4)    & E(CL',LL') & (2,4) \\
E(C'L,LL) & (2,4)    & E(C'L,LL') & (2,4)\\
         &          & E(C'L',LL) & (2,4)\\
\hline
E(CL,CL') & (4,4)    & E(CL,C'L') & (4,8)\rule{0pt}{12pt}\\
E(CL,C'L) & (4,16)   & E(CL',CL') & (4,4)\\
         &          & E(C'L,CL') & (4,16)\\
         &          & E(C'L,C'L) & (4,4) \\
\hline
E(CC,LL') & (2,4)    & E(CC,L'L') & (2,4)\rule{0pt}{12pt}\\
E(CC',LL) & (4,4)    & E(CC',LL') & (4,4)   \\
         &          & E(C'C',LL) & (2,4) \\
\hline
E(CC,CL') & (4,4)    & E(CC,C'L') & (4,8)\rule{0pt}{12pt}\\
E(CC,C'L) & (4,32)   & E(CC',CL') & (8,32)\\
E(CC',CL) & (8,32)   & E(CC',C'L) & (8,32)\\
         &          & E(C'C',CL) & (4,8)\\
\hline
E(CC,CC') & (16,64)   & E(CC,C'C') & (2,8)\rule{0pt}{12pt}\\
         &          & E(CC',CC') & (16,64) \\ \hline
\hline
\end{tabular}
\end{center}
\caption{Cluster cases with constraints on the center of the
  variable-radius circle.  $(m,n)$ denotes the equation degree for
  $E=L$ and $E=C$, respectively.  $L'$ denotes a constraint between a
  line and the center of the variable-radius circle; $C'$ denotes a
  constraint between a circle and the center of the variable-radius
  circle.}
\label{tab:VCCcenter}
\end{table}

\section{Spatial Geometric Constraints}\label{sec:C3d}

Compared to constraint problems in the plane, our knowledge of spatial
constraint systems is relatively modest.  The constraint graph
analysis applies with some notable caveats.  For example, Laman's
characterization of rigidity does not apply in 3-space, not even when
restricting to points only, and distances between them; see Section
\ref{sec:3Dgraphs}.  Furthermore, the subsystems isolated by
constraint graph decomposition can be complex, especially if lines are
admitted to the geometric vocabulary.  We illustrate the latter point
with a few examples.

\subsubsection*{Points and Planes}

Points and planes comprise the most elementary vocabulary in spatial
constraint solving.  Both have three degrees of freedom and are dual
of each other.  In analogy to the minimal constraint graph in 2D
(Definition~\ref{def:minimalCG}), a minimal constraint graph in 3D
consists of three elements, points or planes, and three constraints,
forming a triangle.  The initial placement for the four combinations
places the elements in canonical order, planes first, points second.
Table~\ref{tab:first3d} summarizes the method,~\cite{bib:DurandHo00}.
Note that the constraint between two planes is an angle, and the
constraint between two points or a point and a plane is a distance.
The initial placement fails for the exceptional angles $0$ and $\pi$,
as well as for distance 0 between two points.

\begin{table}
\begin{center}
\begin{tabular}{|l|l|}
\hline
\multicolumn{2}{|c|}{Canonical Placement of three Points or Planes\rule{0pt}{12pt}}\\
\hline
\hline
 & $P_1$ placed as the plane $z=0$ \rule{0pt}{12pt}\\
$P_1,P_2,P_3$ & $P_2$ placed to intersect $P_1$ in the $y$-axis \\
 & $P_3$ placed to contain the origin \\
\hline
 & $P_1$ placed as the plane $z=0$ \rule{0pt}{12pt}\\
$P_1,P_2,p_3$ & $P_2$ placed to intersect $P_1$ in the $y$-axis \\
 & $p_3$ is placed on the $xz$-plane \\
\hline
 & $P_1$ placed as the plane $z=0$ \rule{0pt}{12pt}\\
$P_1,p_2,p_3$ & $p_2$ is placed on the positive $z$-axis \\
 & $p_3$ is placed on the $xz$-plane with $z\geq 0$\\
\hline
 & $p_1$ is placed at the origin \rule{0pt}{12pt}\\
$p_1,p_2,p_3$ & $p_2$ is placed on the positive $x$-axis\\
 & $p_3$ is placed on the $xz$-plane with $z\geq 0$\\
\hline
\end{tabular}
\end{center}
\caption{Placement of three entities that are mutually
         constrained. $P$ denotes a plane $p$ a point.}
\label{tab:first3d}
\end{table}

Sequential constructions of points and planes are straightforward.
The locus of a third point $p$, at respective distances from two known points $p_1$ and $p_2$,
is the intersection of two spheres centered at $p_1$ and $p_2$.
It is a circle that is contained in a plane perpendicular to the line through $p_1$ ad $p_2$.
As before, this fact can be used to simplify the algebra.

The simplest, nonsequential constraint system is the octahedron,
consisting of 6 elements and 12 constraints,
\cite{bib:hoffmann95a,bib:hoffmann95b}.  The name derives from the
constraint graph that has the topology of the octahedron.  There are 7
major configurations according to the number of planes.  The
configurations with 5 and with 6 planes are structurally
underconstrained.  Solutions of the octahedron constraint system have been proposed in
\cite{bib:hoffmann95a,bib:hoffmann95b,bib:Michelucci04,bib:DurandHo00}.
The number of distinct solutions is up to 16.

\subsubsection*{Points, Lines and Planes}\label{sec:C3dSeqLine}

A line in 3-space has four degrees of freedom.  Usually lines are
represented with 6 coordinates, using Pl\"{u}cker coordinates, or with
8, using dual quaternions; e.g., \cite{bib:Blaschke}.
Consequently, the implicit relationships
between the coordinates have to be made explicit by additional
equations when solving constraint systems with lines in 3-space.

The sequential construction of lines can be trivial, for instance
determining a line by distance from two intersecting planes.  But it
can also be hard, for instance, when determining a line by distance
from four given points in space.  The latter problem, in geometric
terms, asks for common tangents to four fixed spheres.  This problem
has up to 12 real solutions.  The upper bound of 12 was shown
in~\cite{bib:hoffmann00} using elementary algebra.  The lower bound of 12
was established in~\cite{bib:Mac} with an example.  In contrast to the
restricted problem of only points and planes, this sequential
constraint problem involving lines is therefore much more demanding.

\begin{figure}
\begin{center}
\includegraphics[scale=0.8]{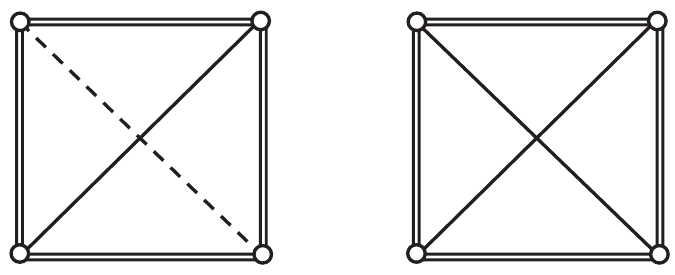}
\end{center}
\caption{The two variational constraint problems with 4 lines.  The
  4 double lines represent both angle and distance, the solid
  diagonals distance, and the dashed diagonal angle constraints.}
\label{fig:4LinSpace}
\end{figure}

We can constrain two lines by distance from each other, by angle, and
also by both distance and angle.  Small constraint configurations that
involve lines have been investigated,~\cite{bib:gao02,bib:gao04b}.
The papers show that there are 2 variational configurations of 4
lines.  They are shown in Figure~\ref{fig:4LinSpace}.  The papers also
show that the number of configurations with 5 geometric elements,
including lines, is 17.  Moreover, when 6 elements are considered, the
number of distinct configurations grows to 683.
Because of this daunting growth pattern, it is natural to seek alternatives.
One such alternative has been proposed in \cite{bib:meiden06,bib:meiden08} where, instead of lines, only segments of lines are allowed.
Consequently, many constraints can be formulated as constraints on end points.
Note that in many applications this is perfectly adequate.

\section{Under-constrained Geometric Constraint Problems}
\label{sec:undercon}

In general, existing geometric constraint solving techniques have been
developed under the assumption that problems are well-constrained,
that is, they adhere to Definition~\ref{def:genwellcon} given in
Section~\ref{sec:intro}.  Put differently, the number of constraints and
their placement on the geometric elements define a problem with a
finite number of solution instances for non-degenerate
configurations.
However, there are a number of scenarios where the assumption of
well-constrained does not apply. Examples are early stages of the
design process when only a few parameters are fixed or in
cooperative design systems where different activities in product
design and manufacture examine different subsets of the information in
the design model, \cite{bib:hoffmannRJA00}.
The problem then is under-constrained, that is,
there are infinitely many solution instances for non-degenerate
configurations.

\begin{figure}
  \begin{center}
  \includegraphics[scale=0.3]{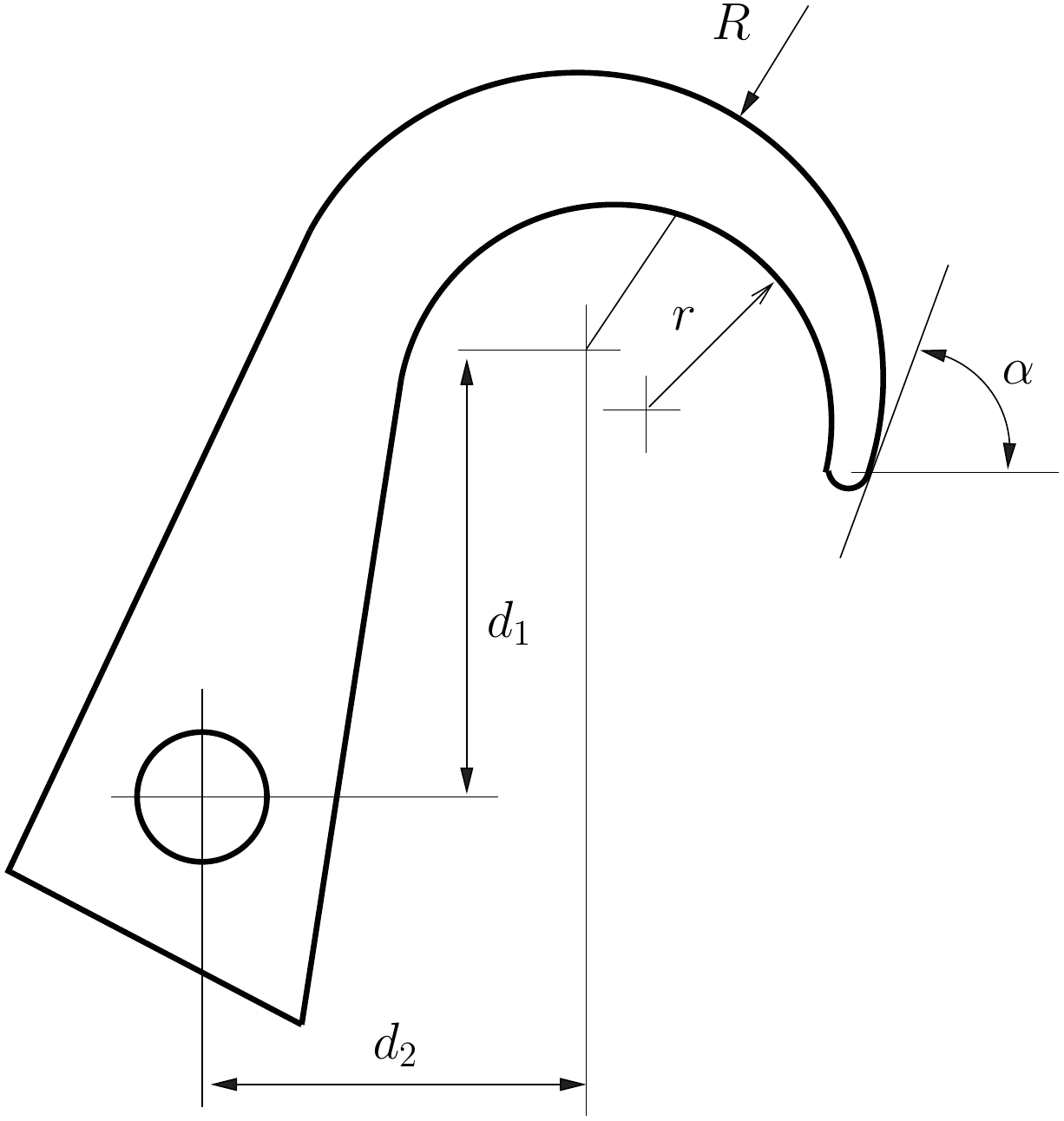}
  \end{center}
  \caption{Hook of a car trunk locker.}
  \label{fig:truncklocker}
\end{figure}

\begin{example}
Consider the hook of a car trunk locker shown in
Figure~\ref{fig:truncklocker}. Once distances $d_1$ and $d_2$ have
placed the center of the exterior circle of the hook with respect to
the hook's axis of rotation, the designer is mainly interested in
finding a value for the angle $\alpha$ where the exterior circle is tangent to the small circle transitioning to the inner circle of the hook.
The angle $\alpha$ has to be such that the hook smoothly rotates while closing and opening the hood.
At this design stage, the stem shape of the hook is irrelevant.
\end{example}

This, and other simple examples taken from Computer-Aided Design,
illustrate the need for efficient and reliable techniques to deal with
under-constrained systems.  The same need is found in other fields
where geometric constraint solving plays a central role, such as
kinematics, dynamic geometry, robotics, as well as molecular modeling
applications.

Recent work on under-constrained GCS with one degree of
freedom,  has brought
significant progress in understanding and formalizing generically
under-constrained systems;
\cite{bib:sitharam10,bib:sitharamUa,bib:sitharamUb}.
The work focuses on GCS restricted to points and distances, also generically called \emph{linkages}.

The  goal of geometric constraint solving is to
effectively determine realizations or embeddings of geometric objects
in the ambient space in which the GCS problem is formulated. Thus, the
current trend is that solving an under-constrained GCS should be
understood as solving some well-constrained GCS derived from the given
one.

There are two ways to transform an under-constrained GCS into a
well-constrained one: adding to
the GCS as many extra constraints as needed or removing from the
GCS unconstrained geometric entities. Note that removing
constrained entities makes little sense.  Accordingly, the
literature on under-constrained
GCS advocates to transform an under-constrained
GCS into a well-constrained one by adding new constraints.
This technique was formally defined in~\cite{bib:joanarinyo03b} as follows.

\begin{definition}\label{def:completion}
Let $G(V,E)$ be an under-constrained graph associated with a GCS problem.
Let $E'$ be a set of additional edges each bounded by two distinct
vertices in $V$ such that the graph $G'(V, E\cup E')$ is
well-constrained. We say that $E'$ is a \emph{completion} for $G$ and
that $G'$ is the \emph{completed} graph of $G$.
\end{definition}

\begin{figure}
  \begin{center}
  \includegraphics[scale=0.7]{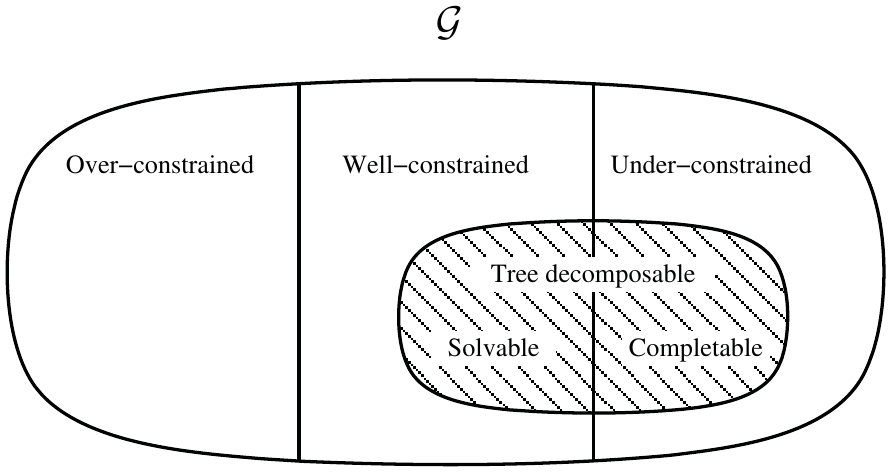}
  \end{center}
  \caption{A partition of the geometric constraint graphs set,
    $\mathcal{G}$. The set of tree-decomposable graphs straddles over
    the sets of well- and over-constraint graphs.}
  \label{fig:Gpartition}
\end{figure}

Let $\mathcal{G}$ denote the set of geometric constraint graphs.
Definitions~\ref{def:genwellcon}, \ref{def:genundercon}
and~\ref{def:genovercon} given in Section~\ref{sec:intro} induce
in $\mathcal{G}$ a partition as shown in Figure~\ref{fig:Gpartition}.
The set of tree-decomposable graphs straddles over the sets of
well- and over-constraint graphs.
As described in Section~\ref{sec:treeDecomp}, the set of
well-constrained, tree-decomposable graphs are solvable
by the tree decomposition approach.

Within the set of under-constrained graphs we can distinguish two families:
Those which are not tree-decomposable and those which are. It is easy
to see that
there is no completion for a graph in the first family that could
transform the graph into a tree-decomposable one.  Considering
graphs in the second family, Definition~\ref{def:genwellcon} fixes the
number of extra constraints that must be added. However deciding which
constraints should actually be added to the graph is not a
straightforward matter because the resulting graph could be either
over-constrained or well-constrained but not tree-decomposable.

\begin{figure}
  \begin{center}
  \mbox{\subfigure[]{\includegraphics[scale=0.5]{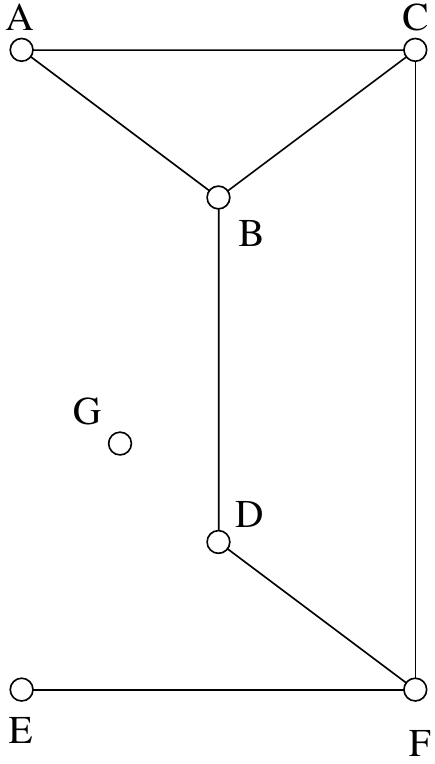}}
        \hspace{1cm}
        \subfigure[]{\includegraphics[scale=0.5]{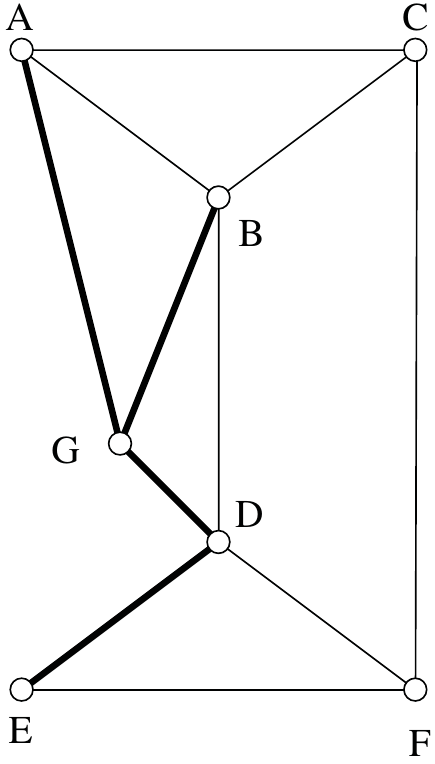}}
        \hspace{1cm}
        \subfigure[]{\includegraphics[scale=0.5]{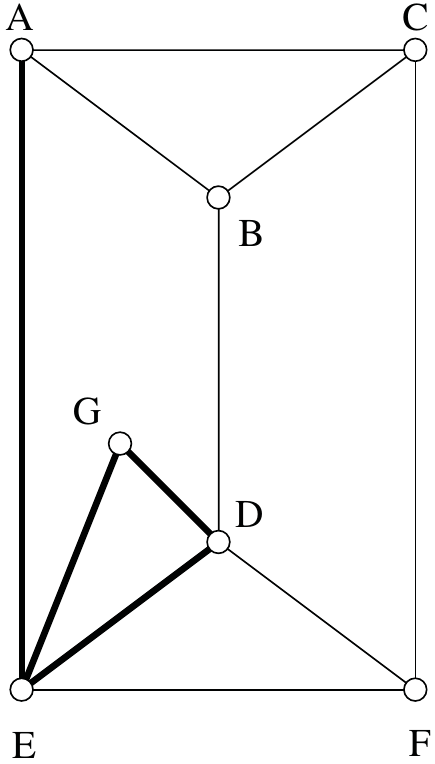}}
  }
  \end{center}
  \caption{a) An under-constrained, tree-decomposable graph $G$.
           b) A tree-decomposable completion of $G$.
           c) A non tree-decomposable completion of $G$.
  }
  \label{fig:twocompletions}
\end{figure}

\begin{example}
Figure~\ref{fig:twocompletions}a shows an under-constrained graph. To
see that it is tree-decomposable just consider as the first
decomposition step the subgraphs induced by the sets of vertices
$\{E,F,G\}$, $\{A,G\}$ and $\{A,B,C,D,F\}$.
Finding the stpdf needed to complete a tree-decomposition is routine.
The completion $E = \{(A,G), (G,B), (G,D), (E,F)\}$ generates the
well-constrained graph shown in Figure~\ref{fig:twocompletions}b.
To see that the graph is tree-decomposable take as a first
decomposition step the two minimal constraint
graphs with edges $\{(E,D)\}$ and $\{(E,F)\}$ plus the subgraph
induced by edges $\{(A,B), (A,C), (B,C), (B,D), (C,F), (D,F)\}$.
However, completion $E=\{(A,E), (E,G), (E,D), (G,D)\}$ results in
the graph depicted in Figure~\ref{fig:twocompletions}c which is no
tree-decomposable.
\end{example}

In what follows we restrict to the completion problem for triangle
decomposable GCS.

\begin{definition}\label{def:completable}
An under-constrained graph $G(V,E)$ is said to be \emph{completable}
if there is a set of edges $E'$ which is a completion for $G$.
\end{definition}

Notice that completability does not require triangle-decomposability
of the underconstrained graph, nor does it imply that a
well-constrained, completed graph be tree-decomposable.

\begin{definition}\label{def:tdcompletion}
Let $G(V,E)$ be a triangle-decomposable, under-constrained graph and
let $E'$ be a completion for $G$. We say that $E'$ is a
\emph{triangle-decomposable completion}, (\emph{td-completion} in
short) for $G$ if $G'(V, E\cup E')$ is triangle-decomposable.
\end{definition}

Reported techniques dealing with under-constrained GCS differ mainly
in the way they figure out completions as well as whether they
aim at figuring out td-completions or just completions.
The work in~\cite{bib:latham96} describes an algorithm where the
constraint graph is captured as a bipartite connectivity graph whose
nodes are either geometric entities or constraints.  Each edge
connects a constraint node with the constrained geometric node.  In
analogy to sequential solvers the graph edges are directed to indicate
which constraints are used to fix (incident) geometric objects.  The
connectivity graph is analyzed according to the degrees of freedom of
under-constrained geometric nodes. Each under-constrained geometric
node is a candidate to support an additional edge to a new constraint,
or if there is an edge that heads a propagation path to an existing
constraint node that has an unused condition.  When there are several
candidates on which the new constraint can be established, the
selection is left to the user. A similar approach to solve
under-constrained GCS based on degrees of freedom analysis is
described in~\cite{bib:noort98}.

Two different notions of td-completion were introduced
in~\cite{bib:joanarinyo03b}. The first one is called
\emph{free completion} and is computed in three
stpdf. First a triangle decomposition for the given graph is figured
out.
Then the set of under-constrained leaf nodes in the decomposition
is identified. Notice that each node in this set stores a subgraph
$G(V,E)$ where $|V| = 2$ and $E$ is the empty set. Thus one edge is
missing.
Finally the completion is computed as the set of missing edges in the
under-constrained leaf nodes of the triangle-decomposition.

\begin{figure}
  \begin{center}
  \mbox{\subfigure[]{\includegraphics[scale=0.5]{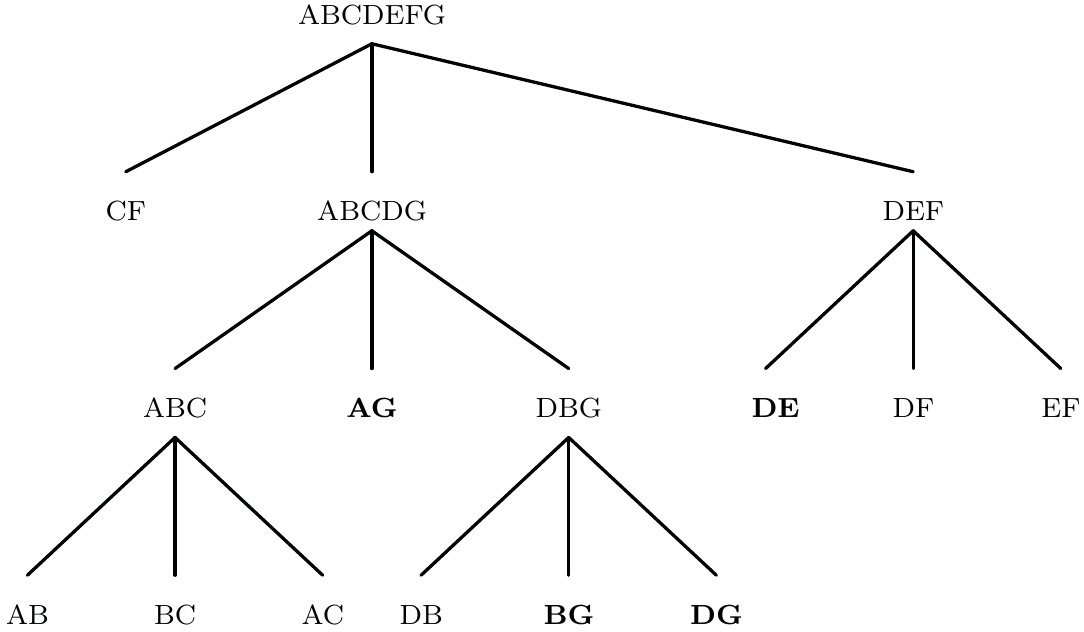}}
        \hspace{1cm}
        \subfigure[]{\includegraphics[scale=0.5]{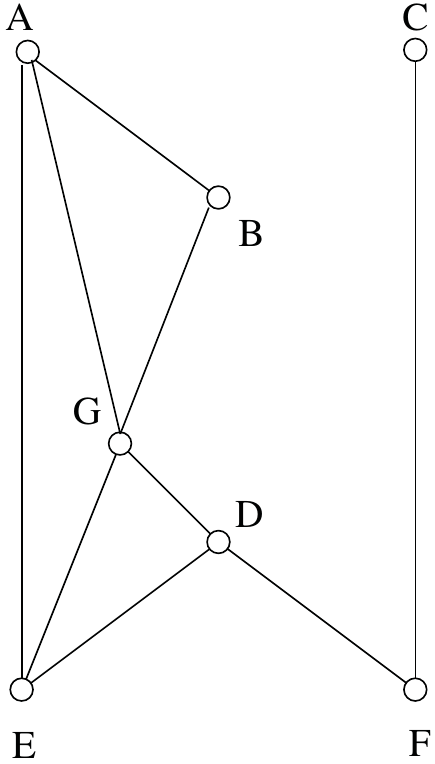}}
        }
  \end{center}
  \caption{a) A triangle-decomposition for the under-constrained graph
              $G(V,E)$ in Figure~\ref{fig:twocompletions}a.
              Pairs of vertices $\{A,G\}, \{D,E\}, \{B,G\}$ and
              $\{D,G\}$ in leaf nodes do not define graph edges.
           b) A set of additional edges defined on $V(G)$. Edges
              $(A,E), (A,G), (B,G), (D,E), (D,G)$ and $(E,G)$
              do not belong to $E$.
  }
  \label{fig:tdunderconsgraph}
\end{figure}

\begin{example}
Figure~\ref{fig:tdunderconsgraph}a shows a triangle-decomposition for
the under-constrained graph $G(V,E)$ in
Figure~\ref{fig:twocompletions}a. We have $|V|=7$ and $|E|=7$.
For 2D problems, the general property of a well-constrained
graph described in Section~\ref{sec:intro} is the Laman
theorem,~\cite{bib:laman70}, $2|V| - |E| = 3$. Hence
the number of additional edges required to complete $G$ to a
well-constrained graph is $|E'| = 2|V|-|E|-3 = 4$. The set of
leaves in the triangle-decomposition corresponding to under-constrained
minimal graphs includes exactly four elements: $\{A,G\}, \{B,G\},
\{D,E\}$ and $\{D,G\}$. Thus $E'=\{(A,G), (B,G), (D,E), (D,G)\}$ is a
free completion for $G$. The completed well-constrained graph
$G'(V,E\cup E')$ is shown in Figure~\ref{fig:twocompletions}b.
\end{example}

The second td-completion is called \emph{conditional completion}. The
first and second stpdf are the same as in the free
completion. However, in the third step, edges to complete
under-constrained leaf nodes in the decomposition are drawn from
an additional graph defined over a subset of vertices of the given
graph. If the number of those edges that are not in the original graph
is smaller than the number required by
Definition~\ref{def:genwellcon}, then the completed graph will remain
under-constrained. However a free completion can eventually be applied
to get a well-constrained completion.

\begin{example}
Consider again the under-constrained graph $G(V,E)$ in
Figure~\ref{fig:twocompletions}a and its triangle-decomposition shown
in Figure~\ref{fig:tdunderconsgraph}a. The set of under-constrained
pairs of vertices in the triangle-decomposition is $\{(A,G),$ $(B,G),$
$(D,E),$ $(D,G)\}$.  Assume that the set of additional edges defined on
$V(G)$ is
$$\{(A,B), (A,E), (A,G), (B,G),(C,F), (D,F), (E,G), (E,D), (G,D)\}$$
as shown in Figure~\ref{fig:tdunderconsgraph}b.
Now, additional edges for the completion must be drawn from a set $E^*$
such that $E \cap E^* = \emptyset$. In the case at hand,
$$E^* = \{(A,E), (A,G), (B,G), (D,E), (D,G), (E,G)\}$$. Thus, a
completion for $G(V,E)$ is
$$E' = \{(A,G), (B,G), (D,E), (D,G)\} \subset E^*$$.
Figure~~\ref{fig:twocompletions}b shows the completed graph
$G'(V,E\cup E')$.
\end{example}

A technique to complete general under-constrained graphs is described
in~\cite{bib:joanarinyo03b}. The approach is based on transforming the
problem of computing a completion into a combinatorial optimization
problem. Edges in the graph and in the additional set are assigned
different weights. Then a greedy algorithm generates a well-constrained
problem provided that there are enough edges in the additional
set. A variant of this approach is reported
in~\cite{bib:zhang06}.

There is a class of 2D, triangle-decomposable, under-constrained GCS that
occur in a number of fields like dynamic geometry or mechanical
computer aided design.  In these GCS the geometries are points, the
constraints are usually point-point distances, and exactly one edge is
missing in the associated constraint graph.  Such GCS are known as
\emph{linkages}.

\begin{figure}
  \begin{center}
    \mbox{\subfigure[]
         {\includegraphics[scale=0.6]{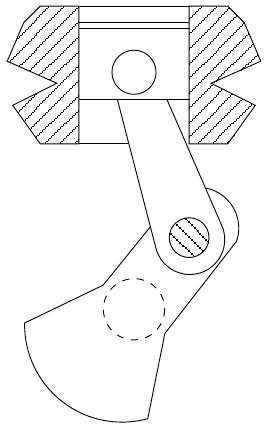}}
         \qquad \qquad
         \subfigure[]
         {\includegraphics[scale=0.7]{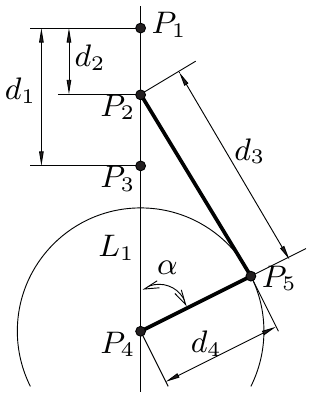}}
         \qquad \qquad
         \subfigure[]
         {\raisebox{0.5cm}
         {\includegraphics[scale=0.5]{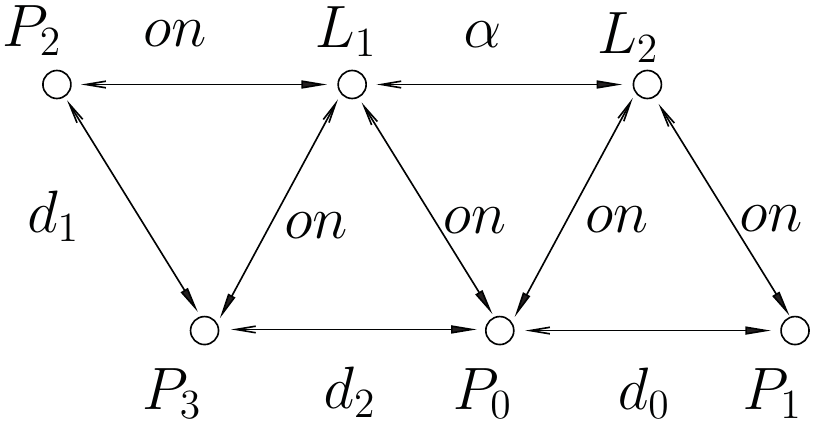}}
         }
    }
  \end{center}
  \caption{a) A crankshaft and connecting rod in a reciprocating
              piston engine.
           b) The crankshaft and connecting rod abstracted as a GCS.
           c) Geometric constraint graph.
  }
  \label{fig:crankshaft}
\end{figure}

\begin{example}
Figure~\ref{fig:crankshaft}a shows an illustration of a crankshaft
and connecting rod in a reciprocating piston engine.
The crankshaft and connecting rod can be abstracted
as the GCS shown in Figure~\ref{fig:crankshaft}b. The GCS
includes four points $P_i$, where $0 \leq i \leq 3$, two lines
$L_1, L_2$, three point-point distances, $d_i, 0 \leq i \leq 2$, one
line-line angle, $\alpha$.  Moreover, the points $P_0, P_2$ and
$P_3$ must be on the line $L_1$, and the points $P_0, P_1$ must be on
the line $L_2$.
If, for example, values of the distance $d_1$ or of the angle $\alpha$
are freely assigned, then the GCS can be considered a linkage.
\end{example}

Reachability is an important problem in fields such as dynamic geometry or
conformational molecular geometry. It can be formalized as follows:
\begin{quote}
Let Rs and Re be two realizations of a well defined geometric
construction where Rs is called the starting instance and Re the
ending instance. Are there continuous transformations that preserve
the incidence relationships established in the geometric construction and
transform Rs to Re?
\end{quote}
The well defined geometric construction in the reachability problem
can be understood as a linkage, that is, a well-constrained GCS
problem where values assigned to one specific constraint can freely
change.  In~\cite{bib:hidalgo13} the reachability problem is solved
assuming that the underlying GCS is triangle-decomposable. The
approach first computes the set of intervals of values that the free
constraint can take for which the linkage is realizable. This set of
intervals is known as the linkage
\emph{Cayley configuration space},~\cite{bib:gao05}.
When both Rs and Re are realizations with the free constraint taking
values within one Cayley interval, the path is an interval arc. When
Rs and Re realizations belong to different intervals, finding a path
entails figuring out whether there are continuous transitions between
consecutive intervals that permit the linkage to reach Re when
starting at Rs.  If more than one such path is found, one is chosen
according to some predefined strategy. In~\cite{bib:hidalgo13} the
minimal arc length path is the one chosen.

\begin{figure}
  \begin{center}
    \mbox{\subfigure[]
         {\raisebox{0.5cm}
             {\includegraphics[scale=0.7]{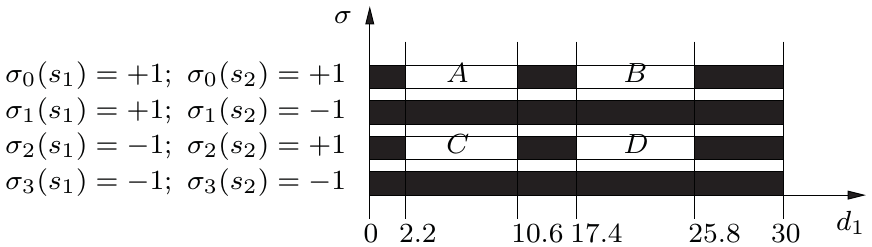}}
         }
         \qquad
         \subfigure[]
         {\includegraphics[scale=0.5]{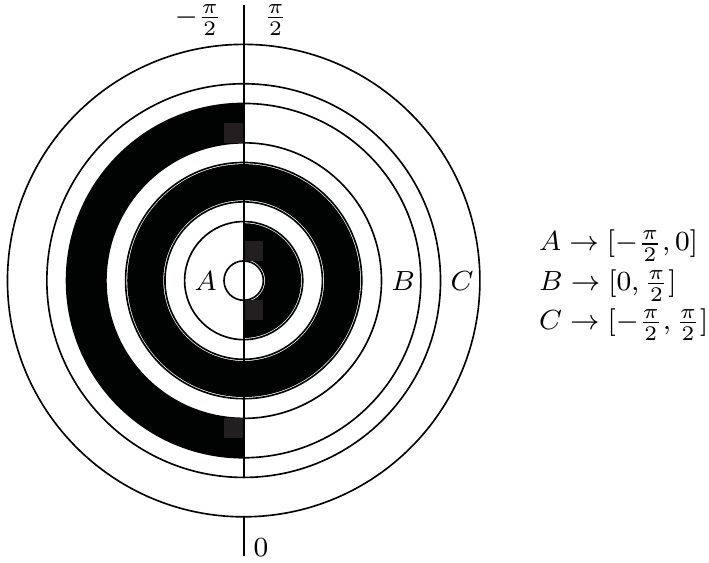}}
    }
  \end{center}
  \caption{a) Cayley configuration for distance constraint $d_2$.
           b) Cayley configuration for angle constraint $\alpha$.
  }
  \label{fig:cayleyConf}
\end{figure}

\begin{example}
  The crankshaft and connecting rod GCS in
  Figure~\ref{fig:crankshaft}b is triangle-decomposable,
  ruler-and-compass solvable. A construction plan that places each
  geometric element with respect to each other is

$$
\begin{array}{llllllll}
1.  &P_0 &= &origin()          &5.  &P_2 &= &intLC(L_1, C_0, s_1)  \\
2.  &P_3 &= &distPP(P_0, d_3)   &6.  &L_2 &= &linePA(P_0, \alpha)   \\
3.  &L_1 &= &line2P(P_0, P_3)   &7.  &C_1 &= &circleCR(P_0, d_0)    \\
4.  &C_0 &= &circleCR(P_3, d_1) &8.  &P_1 &= &intLC(L_2, C_1, s_2) \\
\end{array}
$$
When a construction step has more than one solution, an orientation
parameter is needed to select the desired solution instance. Parameters
$s_1$ and $s_2$ in stpdf 5 and 8 respectively allow to select one of
the two points in a line-circle intersection.
Figure~\ref{fig:cayleyConf}a shows the Cayley configuration space
when the point-point distance constraint value $d_1$ changes.
Intervals labeled $A$, $B$, $C$ and $D$ define orientations and
parameter values for which the GCS is realizable.
Intervals $A$ and $C$ yield realizations consistent
with the one depicted in Figure~\ref{fig:cayleyConf}a. Intervals $B$
and $D$ yield realizations where point $P_2$ would be placed on the line
$L_1$ opposite to $P_1$ with respect to $P_0$.
The Cayley configuration space when the varying parameter is the
angle $\alpha$ is shown in Figure~\ref{fig:cayleyConf}b.
Now orientations are represented along a radial axis and
angle values for which the solution is realizable are represented as
circular intervals.
\end{example}

Linkages are extensively studied
in~\cite{bib:sitharam14,bib:sitharamUa,bib:sitharamUb}. The object of
these works is to lay sound theoretical foundations for a reliable and
efficient computation of Cayley configuration spaces for general
tree decomposable linkages.
New concepts like \emph{size} and \emph{computational complexity} are
introduced and efficient algorithms are developed to answer a
number of questions on linkages like effectively computing Cayley
configuration spaces and solving the reachability problem.
Methods so far applied to compute Cayley configuration spaces, like
the one used in~\cite{bib:hidalgo13}, suffer from potential
combinatorial growth. The work in~\cite{bib:sitharamUa,bib:sitharamUb}
shows that for low Cayley complexity GCS problems,
computing the configuration space is polynomial in the
number of geometric elements of the problem.


\subsection*{Acknowledgement}
Hoffmann gratefully acknowledges partial support by the National Science Foundation under award 1361783.

\bibliography{rja-cons,cmh-extra,if-papers,rja-gen,rja-dyngeom}

\end{document}